\documentclass{aastex63}

\usepackage{newtxtext,newtxmath}

\usepackage[T1]{fontenc}
\usepackage{ae,aecompl}
\usepackage{enumitem}

\usepackage{graphicx}	
\usepackage{amsmath}	
\usepackage{amssymb}	
\usepackage{natbib}

\received{2020 January 10}
\accepted{2020 May 18}

\shorttitle{Significance of Gravitational Nonlinearities}
\shortauthors{Deur, Sargent \& Terzi\'c}

\begin{document}

\title{Significance of Gravitational Nonlinearities on the Dynamics of Disk Galaxies}

\correspondingauthor{Bal{\v s}a Terzi{\'c}}
\email{bterzic@odu.edu}

\author{Alexandre Deur}
\affiliation{Department of Physics, University of Virginia,  \\
Charlottesville, Virginia 22901, USA}

\author{Corey Sargent}
\affiliation{Department of Physics, Old Dominion University,  \\
Norfolk, Virginia 23529, USA}

\author{Bal{\v s}a Terzi{\'c}}
\affiliation{Department of Physics, Old Dominion University,  \\
Norfolk, Virginia 23529, USA}

\begin{abstract}
The discrepancy between the visible mass in galaxies or galaxy clusters, and that inferred 
from their dynamics is well known. The prevailing solution to this problem is dark matter. 
Here we show that a different approach, one that conforms to both the current 
Standard Model of Particle Physics and General Relativity, explains the recently observed tight 
correlation between the galactic baryonic mass and the measured accelerations in the galaxy.
Using direct calculations based on General Relativity's Lagrangian, and parameter-free galactic 
models, we show that the non-linear effects of General Relativity
make baryonic matter alone sufficient to explain this observation.
  Our approach also shows that a specific acceleration scale dynamically emerges. 
It agrees well with the value of the MOND acceleration scale.
\end{abstract}




\section{Introduction}

An empirical tight relation between accelerations calculated from the galactic baryonic content  
and the observed accelerations in galaxies has been 
reported by McGaugh et al.~\citep[hereafter MLS2016]{McGaugh:2016leg}; larger accelerations 
are accounted for by the baryonic matter, i.e.~there is no missing mass problem, while in lower 
acceleration regions, dark matter or gravitation/dynamical laws beyond Newton's are necessary. 
This correlation is surprising because galactic dynamics should be dictated by the total mass 
(believed to be predominantly dark), but instead the baryonic mass information alone is sufficient 
to get the observed acceleration. 
A tight connection between dark and baryonic matter distributions would explain the observation, 
but such connection has not been expected.
While the relation from MLS2016---including its small scatter---can be reproduced with dark 
matter models, \citep{Ludlow:2016qzh}, the consistency of  the measured correlation width with the 
observational uncertainties suggests a dynamical origin rather than an outcome of 
galaxy formation, since this would add an extra component to the width.

Dynamical studies of galaxies typically use Newton's gravity. However, it has been argued
that once galactic masses are considered, relativistic effects arising from large masses 
(rather than large velocities) may become 
important~\citep{Deur_DM-PLB, Deur_DM-EPJC, Deur:2017aas}. Their physical origin is 
that in General Relativity (GR), gravity fields self-interact. In this article, we explore whether these 
effects can explain the relation in MLS2016  without requiring dark matter or modifying gravitation 
as we currently know it.

The article is organized as follows. We first outline the self-interaction effects in GR,
then discuss in Section \ref{depend} the empirical tight dependence of observed acceleration 
on baryonic mass  in disk (i.e.~lenticular and spiral) galaxies. In Section \ref{lattice}, we use 
GR's equations to compute the correlation. 
These CPU-intensive calculations allow us to study only a few galaxies, modeled as 
bulge-less disks. To cover the full range of disk galaxy morphologies, including those 
with significant bulge, in Section \ref{model} we develop two dynamical models of 
disk galaxies in different but complementary ways: uniform sampling (Section \ref{model1}) 
and random sampling (Section \ref{model2}) of the galactic parameter space.
In Section \ref{results} we show the results from these models, and compare them to observations. 
Finally, in Section \ref{conclusion}, we summarize our findings and their importance.
 
\section{Self-Interaction Effects in General Relativity} \label{nonlin}

Field self-interaction makes GR non-linear. The phenomenon is neglected when 
Newton's law of gravity is used, as is typically done in dynamical studies of galaxies or galaxy 
clusters. However, such a phenomenon becomes significant once the masses involved are large 
enough. Furthermore, it is not suppressed by low velocity---unlike some of the more familiar relativistic 
effects---as revealed by e.~g.~the inspection of the post-Newtonian equations~\citep{Einstein:1938yz}. 
In fact, the same phenomenon exists for the strong nuclear interaction and is especially 
prominent for slow-moving quark systems (heavy hadrons), in which case it produces the well-known quark 
confining linear potential. 

The connection between self-interaction and non-linearities is seen 
e.g.~by using the polynomial form of the Einstein-Hilbert Lagrangian \citep[see e.g.][]{Salam,Zee} 
\begin{equation} \label{eq:EH}
\mathcal{L} =  
\frac{\sqrt{\mathrm{det} (g_{\mu\nu})}\, g_{\mu\nu}R^{\mu\nu}}{16\pi G}
=  \sum_{n=0}^{\infty}\left(16\pi GM\right)^{n/2}
\left[\varphi^{n}\big(\partial\varphi\partial\varphi -(16\pi GM)^{1/2}\varphi T\big) \right]
\end{equation}
where $g_{\mu \nu}$ is the metric, $R_{\mu\nu}$ the Ricci tensor, $T_{\mu\nu}$ the energy-momentum tensor, 
$M$ the system mass and 
$G$ is the gravitational constant. In the natural units ($\hbar=c=1$) used throughout this article, 
$[G]=\mbox{energy}^{-2}$. 
The polynomial is obtained by expanding $g_{\mu \nu}$ around a constant metric 
$\eta_{\mu \nu}$ of choice, with 
$\varphi_{\mu \nu} \equiv g_{\mu \nu}-\eta_{\mu \nu}$ the gravitational field. 
The brackets are shorthands for sums over Lorentz-invariant terms~\citep{Deur_DM-EPJC}. 
For example, the $n=0$ term is explicitly given by the Fierz-Pauli Lagrangian~\citep{Fierz:1939ix}:
%
\begin{eqnarray}
\left[\partial\varphi\partial\varphi-\sqrt{16\pi GM}\varphi T\right]=\frac{1}{2}\partial^{\lambda}\varphi_{\mu\nu}\partial_{\lambda}\varphi^{\mu\nu}-\frac{1}{2}\partial^{\lambda}\varphi_{\mu}^{\mu}\partial_{\lambda}\varphi_{\nu}^{\nu}- 
\partial^{\lambda}\varphi_{\lambda\nu}\partial_{\mu}\varphi^{\mu\nu}+\partial^{\nu}\varphi_{\lambda}^{\lambda}\partial^{\mu}\varphi_{\mu\nu}
-\sqrt{(16\pi GM)}\varphi^{\mu\nu} T_{\mu\nu}.
\label{eq:Fierz-Pauli Lagrangian} \nonumber
\end{eqnarray}
 While Eq.~(\ref{eq:EH}) is often used to study quantum gravity---with questions raised 
regarding its applicability in that context, see e.g.~\citep{Padmanabhan:2004xk}---we stress that 
the calculations and results presented here are classical, and thus not subject to the
difficulties arising from quantum gravity nor the issues raised in~\citep{Padmanabhan:2004xk}.
Field self-interaction originates from the $n>0$ terms in Eq.~(\ref{eq:EH}), distinguishing GR 
from Newton's theory, for which the Lagrangian is given by the $n=0$ term. 
One consequence of the $n>0$ terms is that they effectively increase gravity's strength. It is thus reasonable to 
investigate whether they may help to solve the missing mass problem. In fact, it was shown 
that they allow us to quantitatively reproduce the rotation curves of galaxies 
without need for dark matter, also providing a natural explanation for the flatness of the 
rotation curves~\citep{Deur_DM-PLB}. 

The phenomenon underlying these studies is ubiquitous in Quantum Chromodynamics 
(QCD, the gauge theory of the strong interaction). The GR and QCD Lagrangians are similar in that they
both contain field self-interaction terms.  In fact, they are topologically identical 
(see Appendix A where the similarities and differences between 
GR and QCD are discussed). In QCD, the effects of field self-interaction is well-known as
they are magnified by the large QCD coupling, typically $\alpha_{s}\simeq0.1$
at the transition between perturbative and strong regimes~\citep{Deur:2016tte}.
 
In GR, self-interaction effects become important when $GM$---which in the natural unit used in this 
manuscript has a length dimension---reaches a fraction of the characteristic length $L$ of the system. 
Numerical lattice calculations show that $GM \approx 10^{-3}L$ characterizes systems where 
self-interaction cannot be neglected \citep{Deur_DM-EPJC}). At the particle level, gravity, and
 {\it{a fortiori}} its non-linearities, are automatically ignored since $GM_p \approx 10^{-39}r_p$ 
 ($M_p$ and $r_p$ are the proton mass and radius, respectively), and hence 
 $GM_p/r_p/\alpha_s \approx 10^{-40}$. However the ratio becomes $10^{-2}$ for galactic systems, 
 making it reasonable to ask whether QCD-like GR's self-interaction effects should be considered.
That value characterizes e.g.~typical disk galaxies, galaxies interacting in a cluster, and the 
Hulse-Taylor binary. A large mass discrepancy is apparent when the dynamics of galaxies and galaxy 
clusters are analyzed, while the Hulse-Taylor binary is already known to be governed by strong gravity.

In QCD, a critical effect of self-interaction is a stronger binding of quarks, resulting in their 
confinement. 
  In GR, self-interaction likewise increases gravity's binding, which can provide an origin for the 
missing mass problem. However, one may question the relevance of field self-interaction at large galactic 
radii $r$. At these distances the missing mass problem is substantial, while the small matter density should 
make the self-interaction effects negligible. The answer is in the behavior of the gravitational field lines; 
once they are distorted at small $r$ due to the larger matter density, they evidently remain so
even if the matter density becomes negligible (no more field self-interaction, i.e.~no further distortion of the
field lines), preserving a form of potential different to that of Newton. Thus, even if the gravity field becomes 
weak, the deviation from Newton's gravity remains\footnote{An analogous phenomenon exists for QCD: the parton distribution functions (PDFs) that characterize the structure of the proton are non-perturbative objects even if they are defined and measured in the limit of the asymptotic freedom of quarks where $\alpha_s$ tends to zero. Thus, PDFs are entirely determined by the self-interaction/non-linearities of QCD, although those are negligible at the large energy-momentum scale where PDFs are relevant.}.

A key feature for this article is the suppression of self-interaction effects in isotropic and 
homogeneous systems~\citep{Deur_DM-PLB}:
 
\noindent ~~$\bullet$ In a two-point system, large $\sqrt{GM}$ or $\alpha_{s}$ values lead to a constant 
force between the two points (and a vanishing force elsewhere), i.e.~the string-like flux-tube that is 
well-known in QCD. 

 
\noindent  ~~$\bullet$ Due to the symmetry of a homogeneous disk, the flux collapses only outside 
of the disk plane, thereby confining the force to two dimensions. Consequently, the force between the 
disk center and a point in the disk at a distance $r$ decreases as $1/r$.

\noindent ~~$\bullet$ For a homogeneous sphere, the force recovers its 
usual $1/r^2$ behavior since the flux has no particular direction or plane of collapse. 

This symmetry dependence has led to the discovery of a correlation 
between the missing mass of elliptical galaxies and their ellipticity~\citep{Deur:2013baa}. 
This also illustrates the point of the previous paragraph: even if the matter density 
in the disk decreases quickly with $r$, the missing mass problem---which in our approach
comes from the difference between the GR and Newtonian treatments---grows worse since 
the difference between the $ 1/r$ GR force in the 2D disk and the $1/r^2$ Newtonian force grows 
with $r$. 
This offers a simple explanation for the relation reported in MLS2016: although densities, and 
thus accelerations, are largest at small $r$, the $1/r - 1/r^2$ difference between the GR and 
Newtonian treatments remains moderate. However, the difference becomes important at large $r$, 
where accelerations are small. 
Furthermore, at small $r$, the $1/r^2$ force is recovered for GR due to finite disk thickness 
$h_z$, since isotropy is restored for $r \lesssim h_z$.  
This recovery is amplified since disk galaxies often contain a central high-density bulge 
that is usually nearly spherical \citep{MendezAbreu:2007fm}. The departure from the $1/r^2$ 
behavior then occurs after the bulge-disk transition.

\section{Baryonic Mass---Acceleration Dependence} \label{depend}

The correlation between the radial acceleration traced by rotation curves 
($g_{\rm {\scriptscriptstyle obs}}$) and that predicted by the known distribution of baryons 
($g_{\rm {\scriptscriptstyle bar}}$) reported in MLS2016 was 
established after analyzing 2693 points in 153 disk galaxies with varying morphologies, masses, 
sizes, and gas fractions. The MLS2016 authors found a good functional form fitting the correlation:
\begin{equation} \label{eq_gcor}
g_{\rm {\scriptscriptstyle obs}} = 
{{g_{\rm {\scriptscriptstyle bar}}}\over{1-e^{-\sqrt{g_{\rm {\scriptscriptstyle bar}}/g_{\dagger}}}}},
\end{equation}
where $g_{\dagger}$ is an acceleration scale, the only free parameter of the fit. 
In the remainder of the article, we show that the observed correlation may be entirely 
due to the non-linear GR effects which are neglected in the traditional Newtonian analysis. 
In the next section, we use a direct GR calculation of rotation curves for actual 
galaxies modeled as bulge-less disks \citep{Deur_DM-PLB}. We show that when the 
galactic bulge of the actual galaxy is negligible, the calculation yields a relation that agrees with 
the empirical correlation from MLS2016. In the two subsequent sections, we develop models to 
include the effect of bulges and to account for the variation of morphology of disk galaxies.

\section{Direct Calculations} \label{lattice}

The rotation curves of several disk galaxies were computed in \citep{Deur_DM-PLB} based 
on Eq.~(\ref{eq:EH}) and using numerical lattice calculations 
in the static limit~\citep{Deur_DM-EPJC}.  The method  
is summarized in Appendix B. The two-body lattice calculations described there show that given the 
magnitude of galactic masses, the self-interaction traps the field. For a two-body system, 
i.e.~a system characterized by one dominant dimension, field trapping results in a constant force since
a force magnitude at a given distance $r$ is proportional to the field line density crossing an elementary 
surface. Thus, for one-dimensional systems, the force is constant and the potential grows linearly 
with the distance $r$, as obtained in the numerical lattice calculations \citep{Deur_DM-EPJC, Deur_DM-PLB}.
We can extend this result to a two-dimensional system such as a disk. For a field restricted to two dimensions, 
the flux disperses over an angle rather than a solid angle, which yields a force that varies as $1/r$, 
i.e.~obeys a logarithmic potential. Extending the one-dimensional result to the two-dimensional disk 
case of galaxies assumes that the spread of the mass within the disk area does not compromise the 
trapping of the field in two dimensions. This is reasonable since most galactic baryonic mass is 
concentrated near its center. 
This reasoning and the hypothesis that the field remains trapped for a disk are supported by a 
different approach that uses a mean-field method applied to a thin disk distribution~\citep{deur:2020}. 
The mean field calculation yields a large-distance logarithmic potential when the mass of the disk is 
sufficient (see Fig.~\ref{fig:force_bkg} in Appendix B).

\begin{figure}  
\center
\includegraphics[width=0.85\columnwidth]{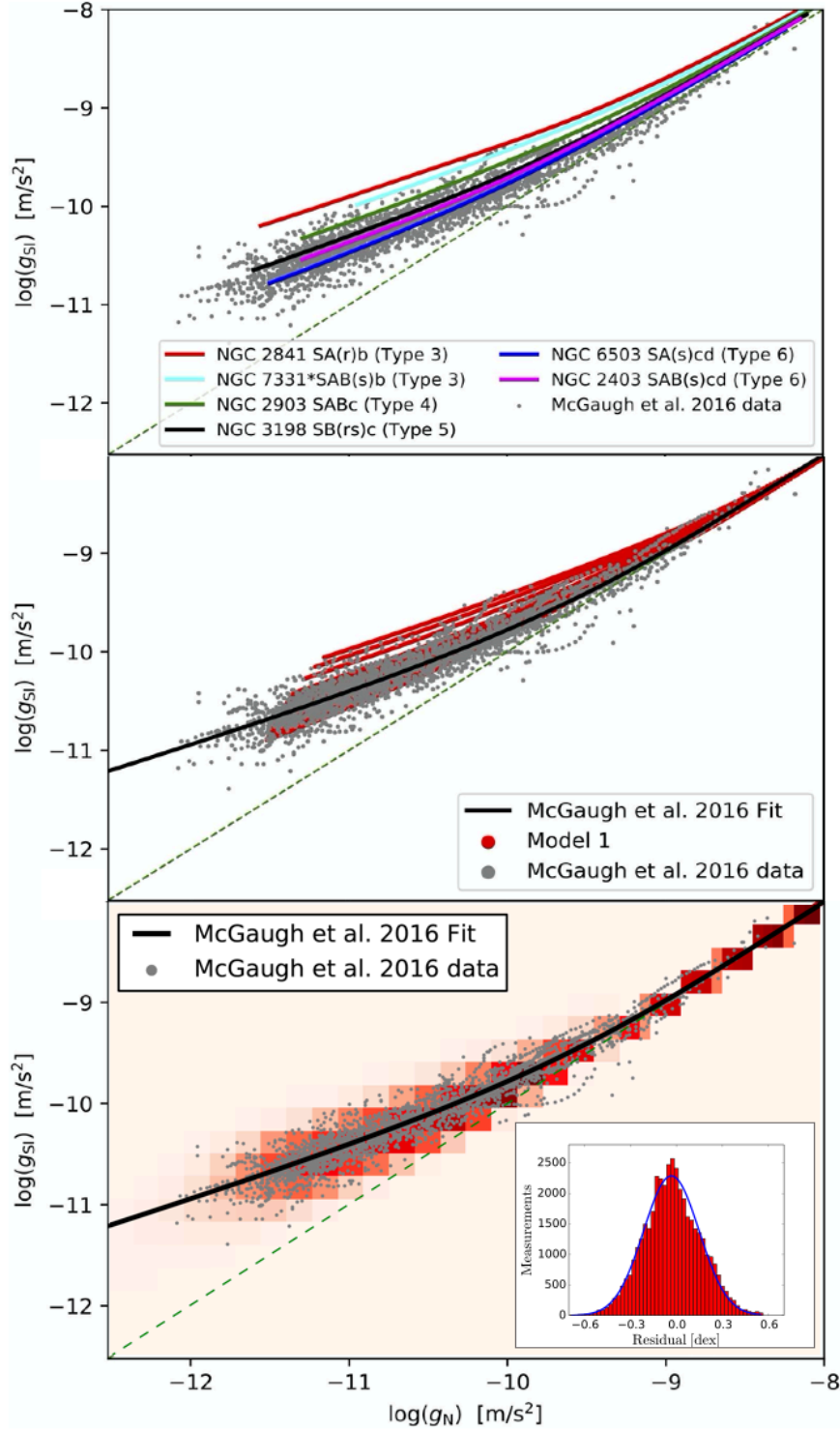}
\vskip-5pt
\caption{Correlation between the acceleration accounting for GR's self-interaction, 
$g_{\rm {\scriptscriptstyle SI}}$, and the acceleration computed with Newtonian gravity,
$g_{\rm {\scriptscriptstyle N}}$, plotted along with the 
 correlation observed in MLS2016 (grey circles). 
Top: Lagrangian-based calculations for various Hubble type galaxies.
The galaxies are approximated as pure (bulge-less) disks. The calculations agree well with 
observation when this approximation is justified (types 5 and 6 galaxies) but depart gradually from 
observation as the bulge becomes more important.
Middle: Model 1 with uniform sampling of the galactic parameter phase space.
Bottom: Model 2 with galactic parameter phase space sampled following observed distributions
(1146 galaxies, sampled at 100 radial values are shown).
The density of the data points obtained with Model 2 is encoded by the color scale.
The dashed line indicates $g_{\rm {\scriptscriptstyle SI}} = g_{\rm {\scriptscriptstyle N}}$.
Embedded in the bottom panel is the residual distribution between the result
of Model 2 and the best fit to the observational data (black line)   beyond the
transition radius, $r>r_t$ (for points within the transition radius, $r\le r_t$, the deviation from
the MLS2016 fit is, by construction, small but systematic; hence the residual has no 
statistical meaning in that region).
}
\label{fig:gcor}
\end{figure}

The calculations of Ref.~\citep{Deur_DM-PLB} neglect the 
galactic bulge and approximate a spiral galaxy with a disk featuring an exponentially-falling 
density profile. They were carried out for nearly bulge-less Hubble types 5 and 6 galaxies 
(NGC 2403, 3198 and 6503), and for Hubble types 3 and 4 galaxies (NGC 2841, 2903 and 7331), 
which have moderate bulges. Using these results, we can compute the total acceleration 
$g_{{\rm {\scriptscriptstyle SI}}}$ stemming from baryonic matter and including GR's field 
self-interaction---analog of $g_{{\rm {\scriptscriptstyle obs}}}$ from MLS2016. 
Plotting it versus the Newtonian acceleration $g_{{\rm {\scriptscriptstyle N}}}$ obtained from the 
same distribution of baryonic matter, but ignoring GR's self-interaction---analog of 
$g_{{\rm {\scriptscriptstyle bar}}}$ from MLS2016---one obtains the results shown 
in the top panel of Fig.~\ref{fig:gcor}. 
%
%
The curves for types 5 and 6 galaxies agree well with the observed correlation, thereby providing 
an explanation for it in bulge-less galaxies. However, the curves for types 3 and 4 galaxies, while 
qualitatively following the correlation, overestimate $g_{{\rm {\scriptscriptstyle SI}}}$ and lie  
on the edge of the observed distribution. That the empirical correlation is reproduced only for 
bulge-less galaxies supports that 1) the correlation from MLS2016 is explainable by 
GR's self-interaction without requiring dark matter  or modification of the known laws of nature, 
and 2) at large acceleration, i.e.~typically
for small galactic radii, the bulge reduces the value of $g_{{\rm {\scriptscriptstyle SI}}}$ since 
self-interaction effects cancel for isotropically distributed matter. 

Although based directly on the GR's Lagrangian, the lattice approach is limited since it is computationally 
costly and applies only to simple geometry, limiting the study to only a few late Hubble type galaxies at 
one time. To study the correlation from MLS2016
over the wide range of disk galaxy morphologies, we developed two models based on: 
1) the $1/r$ gravitational force resulting from solving Eq.~(\ref{eq:EH}) for a disk 
of axisymmetrically distributed matter; and
2) the expectation 
that GR field self-interaction effects cancel for spherically symmetric distributions, such as that of a
bulge, restoring the familiar $1/r^2$ force.

\section{Dynamical Models} \label{model}

To circumvent the limitations of the direct lattice calculation, we constructed two elementary 
models for disk galaxies. They both compute the acceleration including GR's
self-interaction, $g_{\rm {\scriptscriptstyle SI}}$, and the Newtonian acceleration 
due to the baryonic matter, $g_{\rm {\scriptscriptstyle N}}$. 
Both $g_{{\rm {\scriptscriptstyle SI}}}$ and $g_{\rm {\scriptscriptstyle N}}$ are computed 
at a set of radii $r$, from the galactic center to its outermost parts. 
This is carried out for galaxies with their characteristics sampling the observed 
correlations reported in literature. 

The modeled galaxies have two components: a spherical bulge and a larger disk. 
Both contain only baryonic matter following the light distribution, i.e.~there is no dark matter 
and gas is either neglected or follow the stellar distribution.

The bulge is modeled with the projected surface brightness S\'ersic 
profile~\citep{Sersic} used in~\citep{MendezAbreu:2007fm}: 
$I_b(R)=I_{e} 10^{-b_n[(R/R_e)^{1/n}-1]}$, where $R$ is the projected radius, $I_{e}$
is the surface brightness at the half-light radius $R_e$, $n$ is the S\'ersic 
parameter and $b_n \approx 0.868n-0.142$~\citep{caon1993}. 
The internal mass density $\rho_b(r)$, where $r$ is the deprojected radius, 
is computed from the surface brightness by numerically solving the Abel integral.
Since GR's self-interaction effects cancel for isotropic 
homogeneous distributions, the potential in the bulge has the usual Newtonian form, 
$\Phi_b(r)=GM_b^{\rm enc}(r)/r$, where $M_b^{\rm enc}(r)$ is the bulge mass
enclosed within a sphere of radius $r$.

The disk is modeled with the usual surface brightness radial profile $I_d(R)=I_{0} e^{-R/h}$, 
where $I_0$ is the central surface brightness, $h$ is the disk scale length, and possible effects 
from the disk thickness are neglected. 
Again, the corresponding mass density $\rho_d(r)$ is computed from the Abel integral.
Self-interaction in a homogeneous disk leads to a potential $\Phi_d(r)=G'M_d^{\rm enc}(r) \ln (r)$, 
with $M_d^{\rm enc}(r)$ the disk mass enclosed within a radius $r$, and $G'$ the effective 
coupling of gravity in two dimensions, which depends on the physical characteristics of the 
disk~\citep{Deur_DM-EPJC}; see detailed discussion in  Appendix C.

The quantities characterizing a galaxy---the bulge and disk masses, $M_{b}$, $M_{d}$ (from 
which $\rho_{b,0}$ and $\rho_{d,0}$ are obtained, respectively), $R_e$, $n$ and 
$h$---span their observed ranges for S0 to Sd 
galaxies~\citep{MendezAbreu:2007fm, Graham:2008hn, Sofue:2015tsa}.  
There are known relations between these quantities 
\citep{MendezAbreu:2007fm,Sofue:2015tsa}: 
\begin{align} 
& \log(R_e)   =   0.91(7) \log(h)-0.40(3), \label{eq:corr_A} \\
& \log(n)  =   0.18(5) R_e + 0.38(2), \label{eq:corr_B} \\
& \log(M_d)   =  0.58(32)\log(M_b) + 0.002(79). \label{eq:corr_C}
\end{align}
We use values of $R_e$ and $M_b$ from the ranges of observed values 
to obtain the remaining galactic characteristics---$h$, $n$, $M_d$---through 
Eqs.~(\ref{eq:corr_A})-(\ref{eq:corr_C}). Thus, there are no adjustable parameters in our models.

We stress that the accuracy of the empirical relations 
Eqs.~(\ref{eq:corr_A})-(\ref{eq:corr_C}) is not critical to this work, their purpose being only to provide 
reasonable values of the galactic parameter space we select. While the simplicity of our models 
would make it of limited interest for investigating the intricate 
peculiarities of galaxies, such simplicity is beneficial for the present study: no numerous parameters 
nor phenomena (e.g.~baryonic feedback) are needed for adjustment to reproduce the correlation
from MLS2016. That the correlation emerges directly from basic models underlines the fundamental
nature of the correlation.

The two dynamical models introduced in the remainder of this section share the above description. 
From here, they differ in two aspects. The first is in how the observed correlations in 
Eqs.~(\ref{eq:corr_A})-(\ref{eq:corr_C}) are implemented: Model 1 enforces the correlations 
strictly, while Model 2 allows for the parameter space to be randomly sampled. The second 
difference is in representing the transition  radius, $r_t$, between the bulge-dominated 
regime near the center and disk-dominated regime : Model 1 explicitly sets the transition 
at twice the typical bulge scale, $r_t = 2R_e$, while Model 2 defines $r_t$ as the radius
at which the forces due to the two components---the bulge and the disk---are equal.

\subsection{Model 1: Uniform Sampling of the Galactic Parameter Space} \label{model1}

This model generates a galaxy set representative of disk galaxy morphologies by 
uniformly sampling the values of the galactic characteristics discussed in the previous section. 
The model strictly enforces Eqs.~(\ref{eq:corr_A})-(\ref{eq:corr_C}). 
This offers the advantage of simplicity, e.g.~clarity, speed and robustness. 
Actual correlations, however, vary in their strengths. Hence, strictly implementing a 
correlation between quantities $a$ and $b$, and another between $a$ and $c$, would 
result in quantities $b$ and $c$ being also correlated, while if the actual correlations 
between $a$ and $b$, and $a$ and $c$ are both weak, then $b$ and $c$ may not be 
correlated. For example, propagating correlations among different galactic characteristics 
yields an inadequate relation between $M_b$ and $M_d$: 
$M_d \propto M_b^{-\alpha \pm \Delta}$, with $\Delta \gg \alpha=31$.  
To circumvent this problem, we use $h \propto R_e^{1.0}$, 
$M_b \propto R_e^{0.1}$, $n \propto R_e^{0.1}$ and 
$M_d \propto h^{1.0}$, in rough agreement with the correlations from 
Refs.~\citep{MendezAbreu:2007fm, Khosroshahi:1999me}. 

The correlations are applied strictly, i.e.~without accounting for the scatter seen in actual data,
since systematically spanning the observed typical ranges for the quantities contributes to 
the width of the correlation reported in MLS2016, and accounting for such 
scatter would partly double-count, and thus overestimate, the width. 

Inside the spherical bulge-dominated region (denoted by subscript $r<r_t$), 
the self-interaction cancels, and the GR and Newtonian accelerations are the same:
\begin{equation} \label{eq:gs1}
g_{{\rm {\scriptscriptstyle SI}}, r<r_t} (r) = g_{{\rm {\scriptscriptstyle N}}, r<r_t} (r) =
 {{G}\over{r^2}} \left(M^{\rm enc}_{b}(r) + M^{\rm enc}_{d}(r)\right),
\end{equation}
where 
\begin{align}
M^{\rm enc}_{b}(r) & = 4 \pi \int_0^r {\tilde r}^2 \rho_{b}({\tilde r}) d{\tilde r}, \\
M^{\rm enc}_{d}(r) & = 2 \pi \int_0^r {\tilde r} \rho_{d}({\tilde r}) d{\tilde r}.
\end{align}
In the disk-dominated region (denoted by subscript $r>r_t$), 
numerical lattice calculations indicate that self-interaction leads to a collapse in the gravitational 
field lines~\citep{Deur_DM-PLB,Deur_DM-EPJC}. 
The bulge density there is less significant than that of the disk, but is still present. 
The total acceleration is:
\begin{equation} \label{eq:gSI}
g_{{\rm {\scriptscriptstyle SI}}, r>r_t} (r) = {{G}\over{r^2}} M^{\rm enc}_{b}(r) 
+ {{G}\over{r}}' M^{\rm enc}_{d}(r). 
\end{equation}
The Newtonian acceleration $g_{{\rm {\scriptscriptstyle N}}, r>r_t}$ in this region retains the form 
given in Eq.~(\ref{eq:gs1}).

$G'$ is determined by requiring the accelerations to match at $r_t$: 
$g_{{\rm {\scriptscriptstyle SI}}, r<r_t} (r_t) = g_{{\rm {\scriptscriptstyle SI}}, r>r_t} (r_t)$.
Thus, $G' = G/r_t$, by construction. The justification for this choice of $G'$ is explained
in detail in  Appendix C.

The accelerations in the bulge and disk regions are smoothly connected using a 
Fermi-Dirac function centered at $r_t = 2R_e$ and of width $r_t/2$: 
$D(r) =  1/\big(1+ e^{2(r-r_t)/r_t}\big)$. Therefore, the acceleration with self-interaction is: 
\begin{equation}
g_{{\rm {\scriptscriptstyle SI}}}(r) = D(r) g_{{\rm {\scriptscriptstyle SI}}, r<r_t}(r)  
+ \Big(1-D(r) \Big) g_{{\rm {\scriptscriptstyle SI}}, r>r_t}(r), 
\end{equation}
while the Newtonian acceleration is:
\begin{equation} \label{eq:gN}
g_{{\rm {\scriptscriptstyle N}}} (r) = {{G}\over{r^2}} \left(M^{\rm enc}_{b}(r) + M^{\rm enc}_{d}(r)\right).
\end{equation}

The choice of width value for $D(r)$ influences little the result: abruptly transitioning between bulge 
and disk, i.e. using a  step-function rather than $D(r)$, yields quantitatively similar results.  
The small dependence on the functional form for the transition is also supported by the agreement 
between Models 1 and 2 which use different methods for the transition, as we discuss next.

\subsection{Model 2: Random Sampling of the Galactic Parameter Space} \label{model2}

\begin{figure}
\center
\includegraphics[width=0.55\columnwidth]{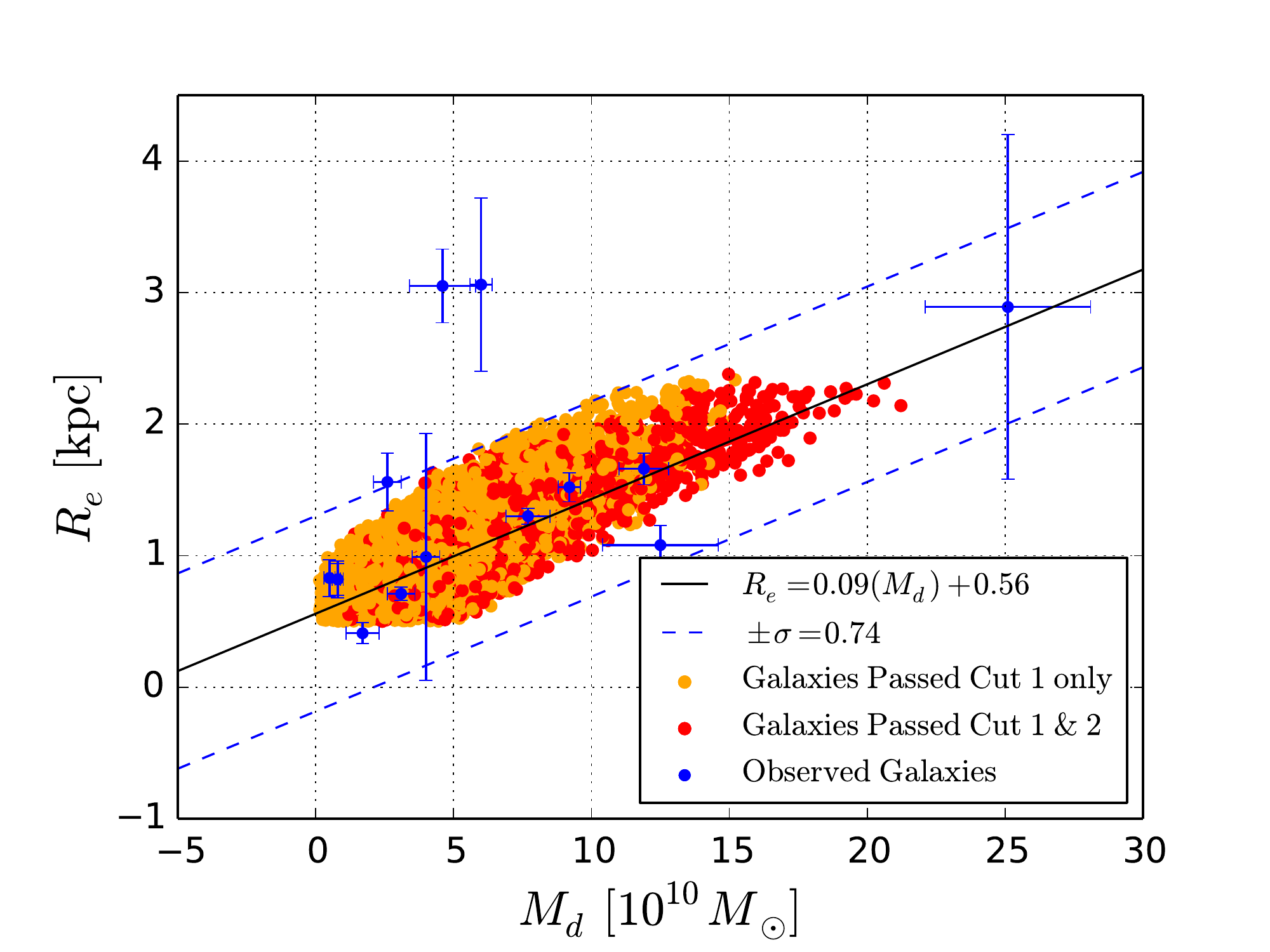}
\caption{$R_e$ vs.~$M_d$. Observed values are shown as blue 
circles~\protect\citep{Sofue:2015tsa}. The best $\chi^2$ fit to the observed
data is denoted with a solid line, and one dex in dashed lines. 
Orange circles denote generated galaxies which passed the first cut only, while the red circles denote 
those who passed both cuts. The galaxies represented by the red circles are disk galaxies, while those
corresponding to the orange circles are too bulge-dominated to qualify as such.}
\label{fig:sample}
\end{figure}

For Model 2, we randomly generate the galaxy characteristics with gaussian distributions centered 
at the observed parameter values, and of widths determined by the observed distributions.
In order to  sample a realistic galaxy parameter space, we apply
two types of cuts on the generated galaxy characteristics. The first type of cut ensures that the randomly 
sampled galaxy characteristics simultaneously satisfy Eqs.~(\ref{eq:corr_A})-(\ref{eq:corr_C}).
 A candidate galaxy is generated by first randomly sampling distributions of $R_e$ and 
$M_b$ separately, and then using them to randomly sample the observed correlations in
Eqs.~(\ref{eq:corr_A})-(\ref{eq:corr_C}) to obtain $h$, $n$ and $M_d$. These are then combined
to find $\rho_{b,0}$ and $\rho_{d,0}$, thereby completing the parameter 
set for a single candidate galaxy. This particular candidate galaxy then passes the first cut if its 
characteristics satisfy all of the correlations to within one standard deviation. 
Galaxies which pass this first cut are shown as orange and red circles in Fig.~\ref{fig:sample}.
The second type of cut is outlined below.

The transition between the bulge-dominated and the disk-dominated regions is implemented with a 
step-function $H(x) = 1$ for $x<0$, and 0 otherwise, such that at $r_t$, 
the acceleration is kept continuous by the proper choice of $G'$. The transition radius $r_t$ is 
defined as the radial location at which the acceleration due to the disk alone is equal to that 
due to the bulge alone:
\begin{equation} \label{eq:rt1}
G{M_b^{\rm enc}(r_t) \over{r_t^2}} = G'{M_d^{\rm enc}(r_t) \over{r_t}}, 
\end{equation}
with $G' = G/r_t$ (see  Appendix C). This choice of $G'$ simplifies the condition for the transition $r_t$ to
\begin{equation} \label{eq:rt2}
M_b^{\rm enc}(r_t) = M_d^{\rm enc}(r_t). 
\end{equation}
Some bulge-dominated galaxies will not have such a transition within $r=100$ kpc, and are 
removed from the sample. This is the second type of cut applied on the parameter space. 
Galaxies that pass both the first and the second types of cuts are shown as red circles in Fig.~\ref{fig:sample}. 
Essentially, orange circles denote galaxies that are largely bulge-dominated and thus cannot qualify
as disk galaxies. The red circles represent those galaxies having small to moderates bulges which qualify 
them as disk galaxies.

Models 1 and 2 use different methods for the bulge-disk transition. The agreement between 
the two models suggests that they are indifferent to a particular method. The acceleration 
including self-interaction is 
\begin{equation} \label{eq:M2}
g_{{\rm {\scriptscriptstyle SI}}}(r) = H(r-r_t) g_{{\rm {\scriptscriptstyle SI}}, r<r_t}(r) 
+ \Big(1-H(r-r_t)\Big) g_{{\rm {\scriptscriptstyle SI}}, r>r_t} (r), 
\end{equation}
where $g_{{\rm {\scriptscriptstyle SI}}, r<r_t}(r)$ and $g_{{\rm {\scriptscriptstyle SI}}, r>r_t}(r)$ are
given in Eqs.~(\ref{eq:gs1})-(\ref{eq:gSI}) and $g_{{\rm {\scriptscriptstyle N}}}(r)$ in Eq.~(\ref{eq:gN}).

In Model 2, we also modeled the effect of the bulge being spheroidal rather than spherical
by introducing a polar dependence: $\rho_b(r,\phi) = \rho_b(r) (1-\epsilon \cos^2\phi)$, with 
$\rho_b(r)$ the spherical bulge density used in Models 1 and 2. This refinement did not noticeably 
change the results, thereby further proving their robustness.

\section{Results} \label{results}

\subsection{Comparison with observations}
\begin{figure}
\center
\includegraphics[width=0.7\columnwidth]{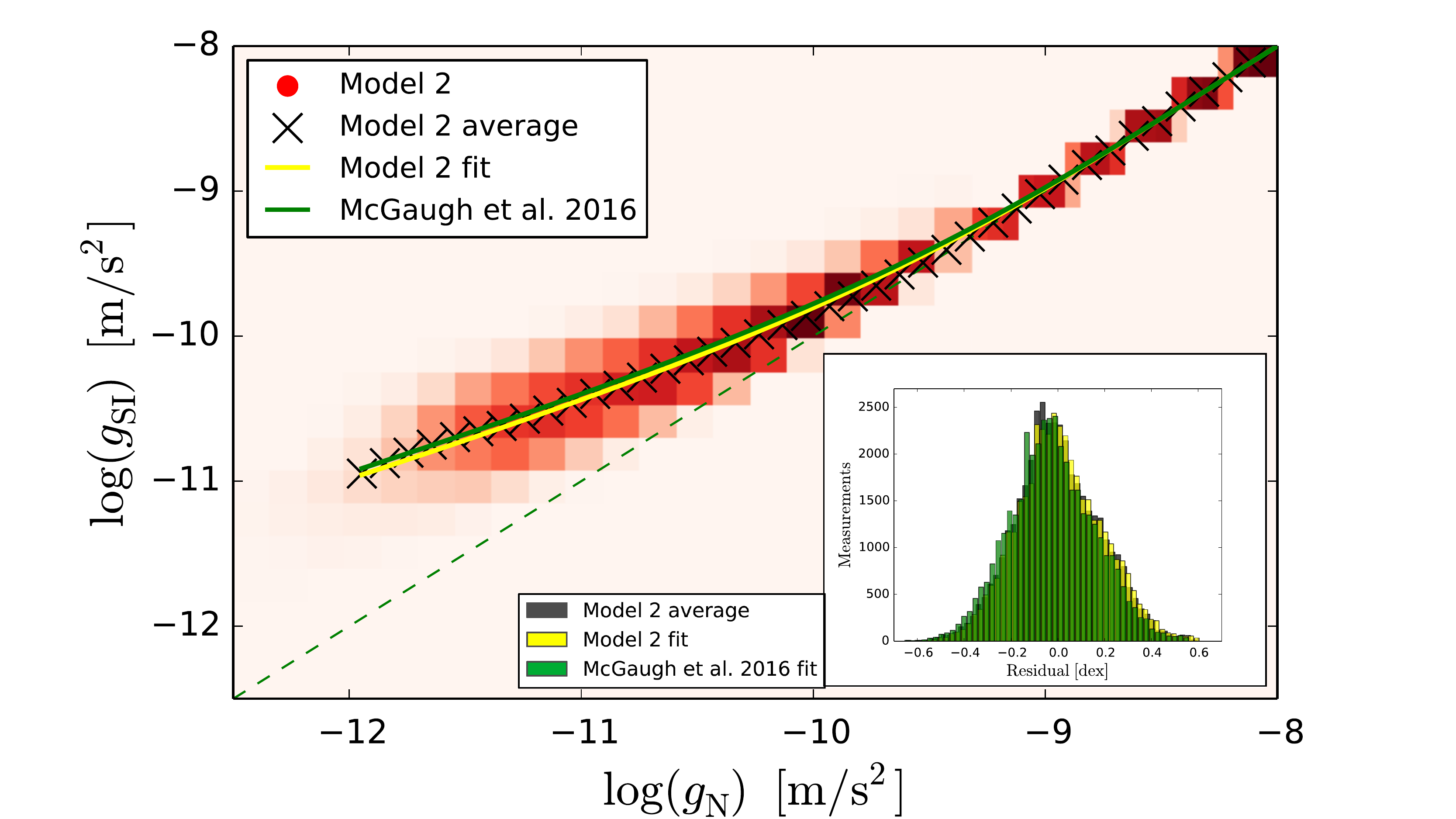}
\caption{
The acceleration accounting for GR's self-interaction, $g_{{\rm {\scriptscriptstyle SI}}}$, versus that 
computed with Newtonian gravity, $g_{{\rm {\scriptscriptstyle N}}}$.
The yellow line shows the best fit to our data simulated with Model 2 (red color density plot) using the 
form in Eq.~(\ref{eq_gcor}). The black x's are the average $\langle \log (g_{{\rm {\scriptscriptstyle SI}}}) \rangle$. 
The yellow line and x's can be compared to the MLS2016 fit, shown by the green line. 
The insert displays the residual between our simulated data and the MLS2016 fit (green histogram, 
already shown in Fig.~\ref{fig:gcor}), the residual using our fit (yellow histogram) and the one using 
$\langle g_{{\rm {\scriptscriptstyle SI}}} \rangle$ (black histogram).}
\label{fig:QualitComp}
\end{figure}

Direct lattice calculation and the two dynamical models allow us to compute the accelerations 
for a set of galaxies whose characteristics follow the typical observed ranges for disk galaxies. 
The acceleration including non-linear self-interaction ($g_{{\rm {\scriptscriptstyle SI}}}$) is plotted  
in Fig.~\ref{fig:gcor} versus the acceleration computed with the same baryonic mass distribution 
but assuming Newtonian gravity ($g_{{\rm {\scriptscriptstyle N}}}$). This is compared to the 
observed correlation between $g_{{\rm {\scriptscriptstyle obs}}}$ and 
$g_{{\rm {\scriptscriptstyle bar}}}$ reported in MLS2016.  
The top panel shows the results for the direct calculation, the middle panel the results for Model 1, 
and bottom panel for Model 2.
Since Model 2 samples the full parameter space selected by the cuts, but with statistical weights favoring 
the more probable parameter space loci, the results must be plotted as data point densities, 
the higher densities being indicated by the darker colors.
Our computed correlations agree well with the empirical observation,  
without invoking dark matter or new laws of gravity/dynamics. 
To quantitatively assess this agreement, we averaged 
$g_{{\rm {\scriptscriptstyle SI}}}$ over all galaxies and also performed a fit of our simulated 
data\footnote{The fit and average 
are performed on the data simulated with Model 2 only. Since
Model 1 samples the galactic phase space uniformly rather than using normal
distributions, a statistical analysis of it would have little meaning.}
 using the same form used in MLS2016, i.e.~Eq.~(\ref{eq_gcor}). The best fit and the average 
$\langle \log (g_{{\rm {\scriptscriptstyle SI}}}) \rangle$ versus $\log(g_{{\rm {\scriptscriptstyle N}}})$ are shown 
in Fig.~\ref{fig:QualitComp}. Our fit parameter  
$g^{\rm Mod2}_{\dagger} = 9.71 \pm 0.27 \times 10^{-11}~{\rm m/s^2}$ is compatible with that
of MLS2016, 
$g_{\dagger} = 1.20 \pm 0.02 \mbox{(stat)} \pm 0.24 \mbox{(syst)} \times 10^{-10}~{\rm m/s^2}$.
This consistency is also manifests in the nearly overlapping residuals displayed in the insert of 
Fig.~\ref{fig:QualitComp}. 
This demonstrates quantitatively the agreement between our model and the data reported in MLS2016.
We must remark that, despite this good agreement, $g^{\rm Mod2}_{\dagger}$ 
was not optimized to fit those of MLS2016,
but that it results directly from Model 2 as described in Section~\ref{model2}.
In fact, it cannot be adjusted since our models have no free parameters. 

%
%
%

\subsection{Emerging characteristic acceleration scale}

\begin{figure}
\center
\includegraphics[width=0.6\columnwidth]{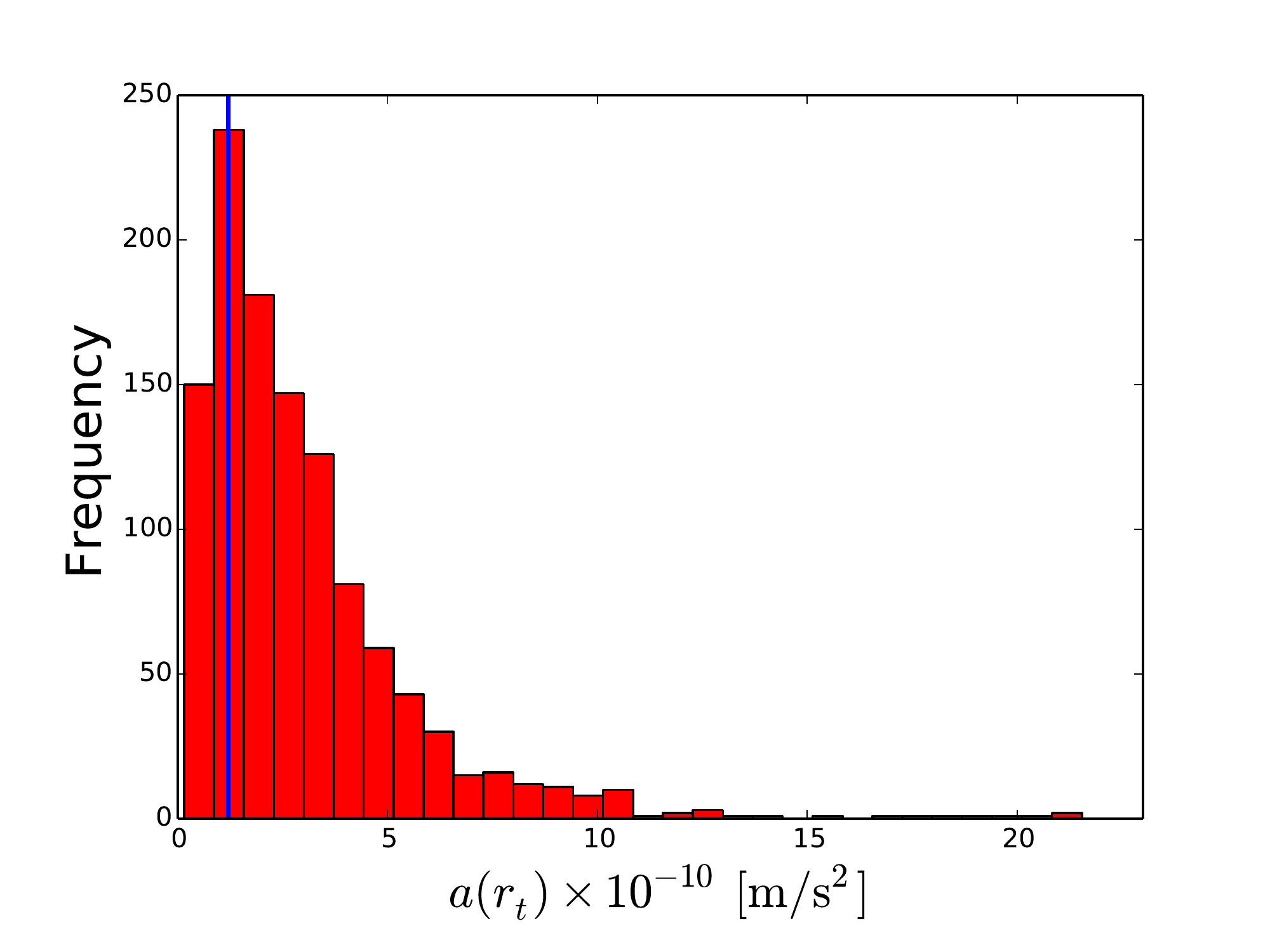}
\caption{Acceleration at the transition radius $r_t$ for the set of galaxies generated in 
Model 2. 
The vertical line denotes $a_0 = 1.2\times 10^{-10}~{\rm m/s^{2}}$ 
from MOND. }
\label{fig:a0}
\end{figure}

The transition scale between the two regimes in Model 2 is defined such as the location where 
forces from the bulge and disk are equal; see Eq.~(\ref{eq:rt1}). 
The acceleration at the transition is shown in Fig.~\ref{fig:a0}, in which the distribution peaks at
$a(r_t)=1.25\pm0.06\times10^{-10}$ ms$^{-2}$.
The sharp peaking indicates that its mode can define a characteristic transition acceleration.       
In our Model 2, this one is consistent with the acceleration parameter $a_0\approx1.2\times10^{-10}$ ms$^{-2}$ 
in the MOND theory \citep{MOND}.
Thus, $a_0$ can be explained as the acceleration at the radius where the self-interaction 
effects become important, that is, in the context of our present model, where the disk mass overtakes the 
bulge mass and causes a transition from the $1/r^2$ 3D force to the $1/r$ 2D force.
For bulgeless disk galaxies, $r_t$ emerges dynamically, see discussion in 
Appendix C, and a direct calculation is necessary to obtain it. 

\subsection{Systematic studies of the residual width}

\begin{figure}
\center
\includegraphics[width=0.5\textwidth]{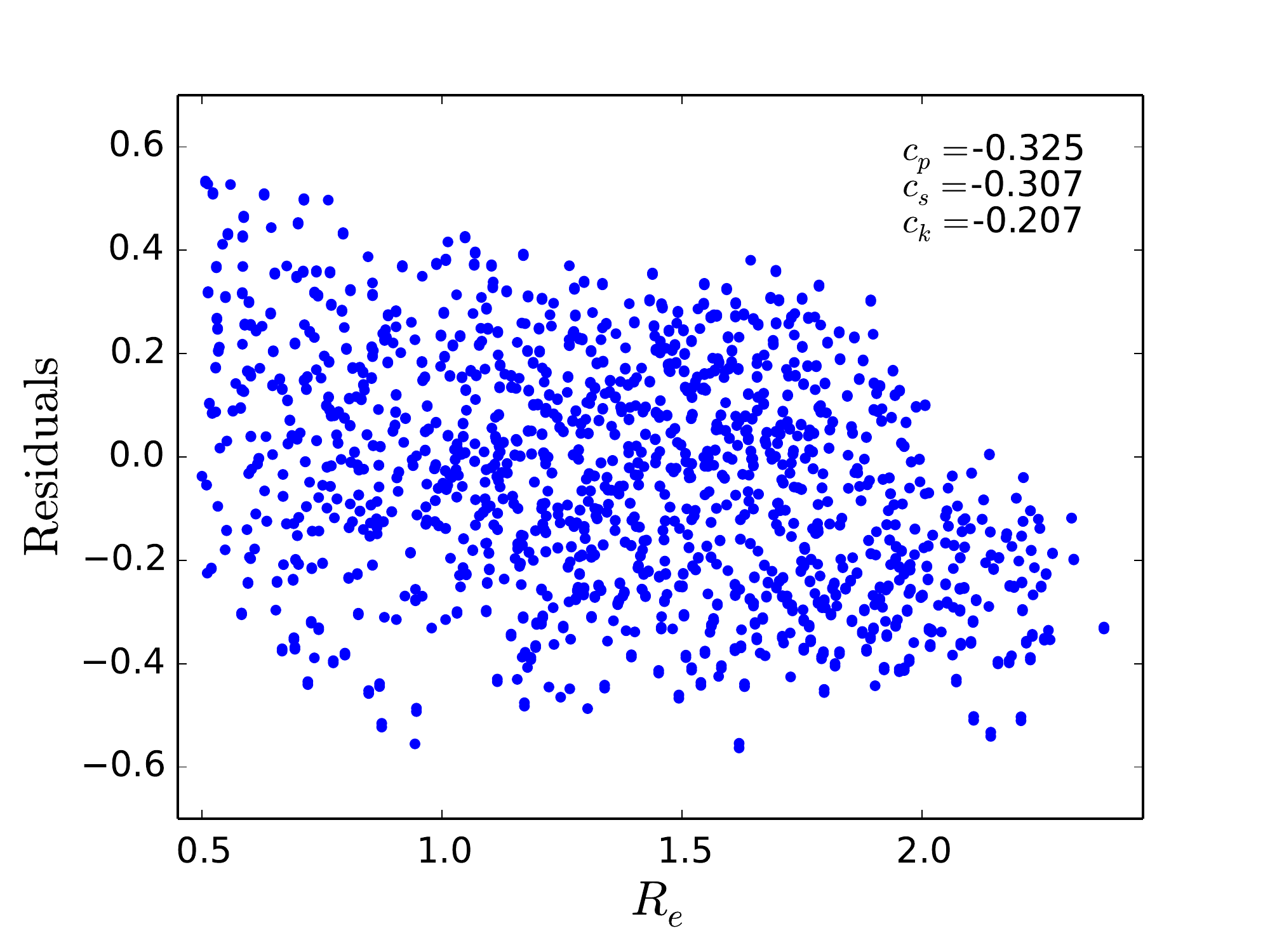} 
\caption{The bulge radius $R_e$ versus the residual between our calculated 
acceleration $g_{{\rm {\scriptscriptstyle SI}}}$ and the MLS2016 relation, shown
for $\log(g_{{\rm {\scriptscriptstyle N}}}) \approx -11$. 
The Pearson's ($c_p=-0.325$), Spearman's ($c_s=-0.307$) and 
Kendall's ($c_k = 0.207$) correlation parameters, all with negligible p-values, indicate 
small correlations. Although they are clear, the correlations do not contribute 
significantly to the residual width. }
\label{fig_correlations}
\end{figure}

One important finding of MLS2016 is that the width of their observed correlation is compatible with 
the uncertainty on the data. This poses a problem for a natural dark matter explanation since the 
baryonic matter-dark matter feedback mechanisms that would be necessary to correlate baryonic 
and dark matter distributions would partly depend on the history of the galaxy 
formation, as shown in MLS2016. 
In the present approach, because of the dependence of $r_t$ on the geometry and mass distributions, 
it may seem at first that the $g_{{\rm {\scriptscriptstyle N}}}$ vs $g_{{\rm {\scriptscriptstyle SI}}}$ correlation 
should depend on the specifics of a particular galaxy, increasing the width of the correlation. 
The criterium for determining $r_t$ in Model 2 is the equality of the disk and bulge forces, and therefore of the accelerations. 
The bulge and disk mass distributions and characteristic lengths being correlated, the acceleration at 
$r_t$ tends to cluster around a single value (see Fig.~\ref{fig:a0}). Because the $g_{{\rm {\scriptscriptstyle N}}}$ 
vs $g_{{\rm {\scriptscriptstyle SI}}}$ correlation is not sensitive to small variations of where the acceleration 
transition happens on the $g_{{\rm {\scriptscriptstyle N}}}=g_{{\rm {\scriptscriptstyle SI}}}$ dashed line of 
Fig.~\ref{fig:gcor}, any dependence on galaxy specificities 
is suppressed. In fact, a change of $a_0$ by the variance extracted from Fig.~\ref{fig:a0} does not 
appreciably affect Fig.~\ref{fig:gcor}.
We can quantitatively verify this by investigating whether large correlations exist 
between the galaxy characteristics and the residual shown in Fig.~\ref{fig:gcor}.
Large correlations
would disagree with the MLS2016 finding 
that their relation has no intrinsic width
, and with the further verification in Ref.~\citep{Lelli_2017}
that the MLS2016 residual does not correlate with galaxy properties.  
We used the Pearson correlation coefficient $c_p$ to check for linear 
correlations between the residual and each galaxy properties---$R_e$, $h$, $M_b$ and $M_d$. 
Since the possible correlations could be non-linear, we also used the Spearman $c_s$ and 
the Kendall $c_k$ rank correlation coefficients. 
To maximize the sensitivity, we investigated the 
correlations at a fixed acceleration value, selected to be $-11.1 \leq \log(g_{{\rm {\scriptscriptstyle N}}}) \leq -11$, {\it{viz}} we checked whether 
galaxy characteristics are correlated with $g_{{\rm {\scriptscriptstyle SI}}}$, along the vertical line at  
$\log(g_{{\rm {\scriptscriptstyle N}}}) \approx -11$ for the simulated data shown on the bottom panel of Fig.~\ref{fig:gcor}. 
The value $\log(g_{{\rm {\scriptscriptstyle N}}})  \approx -11$ is optimal because there, the width is large, 
which maximizes the sensitivity to possible correlations, while the statistics remain important. 
Selecting $\log(g_{{\rm {\scriptscriptstyle N}}}) \approx -11$ and 
computing correlation coefficients reveals that small correlations are present between the residual and the 
galaxy characteristics, see Fig.~\ref{fig_correlations}
for an example with $R_e$. (The other galaxy characteristics $h$ and $M_b$ also display correlations, albeit smaller. $M_d$ is not correlated.)
To quantitatively investigate the effect of these correlations on the
$g_{{\rm {\scriptscriptstyle N}}}$ vs $g_{{\rm {\scriptscriptstyle N}}}$ relation, we first take note that 
they are largely linear. 
This is suggested by the value of $c_p$ being similar to $c_s$ and $c_k$, as well as 
by the fact that polynomial fits of the residual vs galaxy characteristic distribution
are numerically close to a linear fit. The approximate linearity of the 
correlations is confirmed by fitting linearly the correlations, then removing the linear dependence 
using the fit result. While $c_p$ calculated for the modified distributions must be zeroed by construction, 
the new $c_s$ and $c_k$ would reveal no remaining correlation only if the initial correlations had been linear. 
We indeed find negligible values of all the correlation coefficients for the modified distributions, e.g. $c_p=-6\times10^{-17}$, 
$c_s=-4\times10^{-3}$ ($p-$value $0.85$) and 
$c_k=-2\times10^{-3}$ ($p-$value $0.87$) for $R_e$. 
By simultaneously applying this procedure for the distributions 
of the residual versus $R_e$, $h$, $M_d$ or $M_b$, we obtain a rms of 0.1812 for the modified
residual distribution. Comparing with the 
rms for the initial residual distribution, 0.2101, we conclude that while the correlations are 
clear, as shown by their correlation coefficients and negligible p-values, their effects on the residual width 
are small, increasing it by 13\%.

%

\section{Discussion and Conclusion} \label{conclusion}

Our findings support the possibility that GR's self-interaction effects increase the gravitational force 
in large, non-isotropic mass distributions. When applied to disk galaxies, the increased 
force on the observed matter transposes to the missing mass needed in the traditional 
Newtonian analyses. We have thus proposed a plausible explanation for the correlation 
between the luminous mass in galaxies and their observed gravitational acceleration shown in 
MLS2016. That this correlation is encapsulated in our models, free of adjustable parameters,
indicates its fundamental origin.   This work also offers a 
possible explanation for the MOND acceleration scale $a_0$, showing that it dynamically emerges 
from galaxy baryonic mass distribution. Thus, in our approach, the emergence of $a_0$ is due to 
complexity, rather than new physics, such as modifying gravity or Newton's dynamical law.

The explanation proposed here is natural in the sense that it is a consequence of the fundamental 
equations of GR and of the characteristic magnitudes of the galactic gravitational fields, and in the 
sense that no fine tuning is necessary. This contrasts with the dark matter approach that necessitates 
both yet unknown particles and a fine tuning in galaxy evolution and baryon-dark matter 
feedbacks~\citep[see e.g.][]{Ludlow:2016qzh}.
We used several approaches that are quite different, thus leading to a robust conclusion.

The work presented here adds to a set of studies that provide straightforward and natural explanations 
for the dynamical observations suggestive of dark matter and dark energy, but without requiring them 
nor modifying the known laws of nature. This includes flat rotation curves of 
galaxies~\citep{Deur_DM-PLB} 
and the evolution of the universe~\citep{Deur:2017aas}. The Tully-Fisher relation~\citep{Tully:1977fu} 
also finds an immediate explanation~\citep{Deur_DM-PLB}.
There are compelling parallels between those observations and QCD 
phenomenology, e.g.~the equivalence between galaxies' Tully-Fisher relation, and hadrons' Regge 
trajectories~\citep{Deur_DM-PLB, Deur_DM-EPJC}, plausibly due to the similarity between 
GR's and QCD's underlying fundamental equations.
The fact that these phenomena are well-known for other 
areas of nature that possess a similar basic formalism; the current absence of natural and compelling 
theory for the origin of dark matter (supersymmetry being now essentially ruled out); 
and the yet unsuccessful direct detection of a dark matter candidate or its production in 
accelerators despite coverage of the phase-space expected for its characteristics; all support 
the approach we present here as a credible solution to the missing mass problem.

\section*{Acknowledgements}

We are grateful to S. McGaugh for kindly sharing the raw data from his publication.
This work is done in part with the support of the U.~S.~National Science Foundation award 
No.~1535641 and No.~1847771. 




\vskip10pt
 
\section*{Appendix A: Parallels between galaxy dynamics, gravitation, and the strong interaction}

Quantum chromodynamics (QCD), the gauge theory of the nuclear strong interaction is the archetype 
of an intrinsic non-linear theory. The non-linearities are intrinsic since they are present even in the pure 
field case, that is when matter is not present. This contrasts with electromagnetism (QED) which is 
linear for pure-field and for which non-linearities appear only when matter fields are present. GR 
possesses the same intrinsic non-linearities as QCD. In fact, the QCD field Lagrangian is topologically
equivalent to that of field part of the GR Lagrangian given in Eq.~(\ref{eq:EH}). 
This is seen by developing the standard expression 
of the QCD Lagrangian density in term of the gluon field strength $F_{\mu \nu}$ as:

\begin{eqnarray}
\label{eq:QCD Lagrangian}
\mathcal{L}_{\rm QCD} 
& = & 
-\frac{1}{4}F_{\mu \nu}^a F^{\mu \nu}_a
= 
\frac{1}{4}\big(\partial_{\nu}A_\mu^a - \partial_{\mu}A^a_\nu  \big) \big(\partial^{\mu}A^{\nu a} 
- \partial^{\nu}A^{\mu a}  \big) 
+
\sqrt{\pi \alpha_s}f^{abc}\big(\partial_{\nu}A_\mu^a - \partial_{\mu}A_\nu^a\big)A^{\mu b}A^{\nu c} \\
& - & \pi \alpha_s f^{abe}f^{cde}A^a_\mu A^b_\nu A^{\mu c} A^{\nu d}
+ {matter~term}, \nonumber 
\end{eqnarray}
with $A_\mu^a$ the gluon field and with the SU(3) color index $a=1,\ldots, 8$. $f^{abc}$ are the SU(3) 
structure constants and $\alpha_s$ is the QCD coupling. The $matter~term$ is the usual Dirac 
Lagrangian with a covariant derivative and color indices. With the bracket short-hand notation used for 
Eq.~(\ref{eq:EH}), which now also includes summation over color indices, the QCD Lagrangian has the form:
\begin{eqnarray}
\label{eq:QCD Lagrangian2}
\mathcal{L}_{\rm QCD} =\big[\partial A \partial A  \big] 
+\sqrt{16\pi \alpha_s}\big[A^2 \partial A\big]
-4\pi \alpha_s \big[A^4 \big]  + \ matter~term. 
\end{eqnarray}
%
As for GR, the first term is the linear part of the theory and the higher terms are the pure 
field self-interaction vertices. 

While GR and QCD have similar underlying fundamental equations for the pure-field 
part of their Lagrangians, they also have important differences: 
\begin{enumerate}[label=\Alph*), noitemsep,topsep=0pt]
\item GR is a classical field theory while QCD is a quantum field theory;
\item The gravitational field is tensorial (spin-2) while QCD's field is vectorial (spin-1). Consequently, 
gravity is always attractive while color charges in QCD can be attracted or repulsed;
\item $G$ is very small ($GM_p^2=5.9\times10^{-39}$, with $M_p$ the proton mass), while $\alpha_s$
 is large ($\alpha_s\approx 0.1$ at the transition between the weak and strong regimes of
 QCD~\citep{Deur:2016tte});
 \end{enumerate}
However, these differences do not invalidate the parallel between QCD and GR in the context of astronomy.
 
Regarding difference A, classical effects are usually associated with Feynman tree diagrams, while 
quantum effects typically emerge from loop diagrams. The latter cause the 
scale-evolution of the field coupling, while the former generate (in particular) field self-interaction. 
Thus, quantum effects are necessary for quark confinement since they cause $\alpha_s$ to increase enough 
so that the confinement regime is reached even with only a few color charges involved. However, the scale-evolution of $\alpha_s$ is not the basic mechanism for 
confinement\footnote{The apparent divergence of $\alpha_s$ at long 
distance due to scale-evolution had lead to an erroneous explanation of QCD's confinement in term the 
force coupling becoming infinite. However, the divergence is an artifact of applying 
a perturbative formalism in a non-perturbative regime, and this explanation for quark confinement is now 
disproven~\citep{Deur:2016tte}}. 
It is the 3-gluon and 4-gluon interaction tree-diagrams that are at its root. 
In fact, bound states of QCD can be described semi-classically as divergences of perturbative 
(i.e.~with a finite and relatively small value of the QCD coupling) expansion in 
ladder-type Feynman diagrams~\citep{Dietrich:2012un}.
Other semi-classical approaches to hadronic structure---which is ruled by QCD---exist, such as 
AdS/QCD~\citep{Brodsky:2014yha}, and reproduce efficiently the strong QCD
phenomenology~\citep{Brodsky:2010ur}. 
To summarize, while quantum effects are necessary to enable the QCD confinement regime, 
the underlying mechanism for confinement is arguably classical. 
GR and QCD Lagrangians have identical tree-diagrams and thus GR, irrespective to its classical nature, 
should also exhibit effects akin to confinement once its effective field coupling $\sqrt{GM}$ is large enough. 
In fact, black holes are fully confining solutions of GR.

Differences B and C essentially compensate each other.
Classically, a force coupling is truly constant and $G$, in contrast to $\alpha_s$, remains small. 
However, a large effective field coupling can still occur since the magnitudes of the fields are themselves large.
They are proportional to $\sqrt{M}$ with $M$ the mass of the field source. 
This ultimately arises from the tensorial nature of the gravity field, which makes it always attractive: 
gravitational effects can cumulate into large $M$, such as those characterizing galaxies. 
This effectively provides the large coupling in lieu of the (quantum) scale-evolution of the coupling. 
Therefore, the differences B and C balance each other for massive enough systems and self-interaction 
effects similar to the ones seen in QCD should also occur in massive gravitational systems.

The analogous form of GR and QCD field Lagrangians, and the fact that large effective couplings are 
possible for both QCD and GR, may explain intriguing similarities between observations 
suggestive of dark matter and dark energy, and the phenomenology of hadronic structure:
\begin{itemize}[noitemsep,topsep=0pt]
\item Just like for hadrons, the total masses of galaxies and galaxy clusters appear much 
larger than the sum of their known constituent masses.

\item Hadrons and galaxies obey similar mass-rotation 
correlations (Regge trajectories~\citep{Regge:1959mz} and the Tully-Fisher relation~\citep{Tully:1977fu}, 
respectively); In both case $J \propto M^\alpha$ with $J$ the angular momentum, $M$ the (baryonic) mass 
of the system and $\alpha=1.26 \pm 0.07$ or $\alpha=2$ for disk galaxies or hadrons, respectively. 
(The difference in the value of $\alpha$ is due to the difference in the system symmetry, 
see \citep{Deur_DM-EPJC}.)

\item The large scale arrangement of galaxies into filaments is reminiscent of QCD strings/flux tubes;

\item The nucleon and galaxy matter density profiles both decrease exponentially;

\item The approximate compensation at large-scale between dark energy and matter's gravitational 
attraction---a phenomenon known as the cosmic coincidence problem---is comparable to the approximate
suppression of the strong force at large-distance, i.e.~outside the hadron~\citep{Deur:2017aas}.
\end{itemize}
These parallels and the similar form of QCD and GR's Lagrangians suggest that very massive structures such
as galaxies or cluster of galaxies have entered the non-linear regime of GR, and that phenomena linked to 
the dark universe may be the consequence of neglecting this regime.

~

\section*{Appendix B: Summary of the method used in the direct calculations}

The direct calculation of the effects of field self-interaction based on Eq.~(\ref{eq:EH}) employs the Feynman 
path integral formalism solved numerically on a lattice.
While the method hails from quantum field theory, it is applied in the classical limit, see~\citep{Deur_DM-EPJC}. 
The first and main step is the calculation of the potential between two essentially static ($v \ll c$) sources in the 
non-perturbative regime. Following the foremost non-perturbative method used in QCD, we employ a lattice 
technique using the Metropolis algorithm, a standard Monte-Carlo 
method~\citep{Deur_DM-PLB, Deur_DM-EPJC}. The static calculations are performed on a 3-dimensional 
space lattice (in contrast to the usual 4-dimensional Euclidian spacetime lattice 
of QCD) using the 00 component of the gravitational field $\varphi_{\mu \nu}$. This implies that the results 
are taken to their classic limit, as it will be explained below. 
Furthermore, the dominance of $\varphi_{00}$ over the other components of the gravitational field 
simplifies Eq~(\ref{eq:EH}) in which 
$\left[\varphi^{n}\partial\varphi\partial\varphi \right] \to a_n \varphi_{00}^{n}\partial\varphi_{00}\partial\varphi_{00} $, 
with $a_n$ a set of proportionality constants. One has $a_0\equiv 1$ and one can show that $a_1=1$~\citep{Deur_DM-EPJC}.

In this Appendix, we denote $\varphi\equiv\varphi_{00}$ and we will explicitly write $\hbar$ in the expressions in order to identify the quantum effects.

The instantaneous potential from a point-like source located at $x_1$ is given at location $x_2$ by the 
two-point Green function $G_{2p}(x_1-x_2)$. In the path-integral formalism,
\begin{equation} \label{eq:2pt green}
G_{2p}(x_{1}-x_{2})=\frac{1}{Z}\intop\mathrm{D}\varphi\,\varphi(x_{1})\varphi(x_{2})\mathrm{e}^{-\mathrm{i}\, \frac{S_{\mathrm{s}}}{\hbar}},
\end{equation}
with $S_{\mathrm{s}}\equiv{\int{\mathrm{d}^{4}x\,\mathcal{L}}}$ the action, 
$Z\equiv\intop{\mathrm{D}\varphi\,\mathrm{e}^{-\mathrm{i}\, \frac{S_{\mathrm{s}}}{\hbar}}}$, and 
$\intop{\mathrm{D}\varphi}$ is the sum over all possible field configurations. In the lattice method, $Z\equiv1$.
For Euclidian spacetime lattice simulations, one dimension is the time direction. Suppressing it by considering 
static or stationary systems allows us to identify $G_{2p}$ to the instantaneous potential. In that case, the sum 
$\intop{\mathrm{D}\varphi}$ is over configurations in position space only. 
This allows us to perform standard lattice calculations of difficult forces such as gravity, in spite of its 
tensorial nature. The method described in the next paragraph is thus the standard 
one described in lattice textbooks.

\begin{figure}[ht!] 
\centering
 \includegraphics[width=0.45\textwidth]{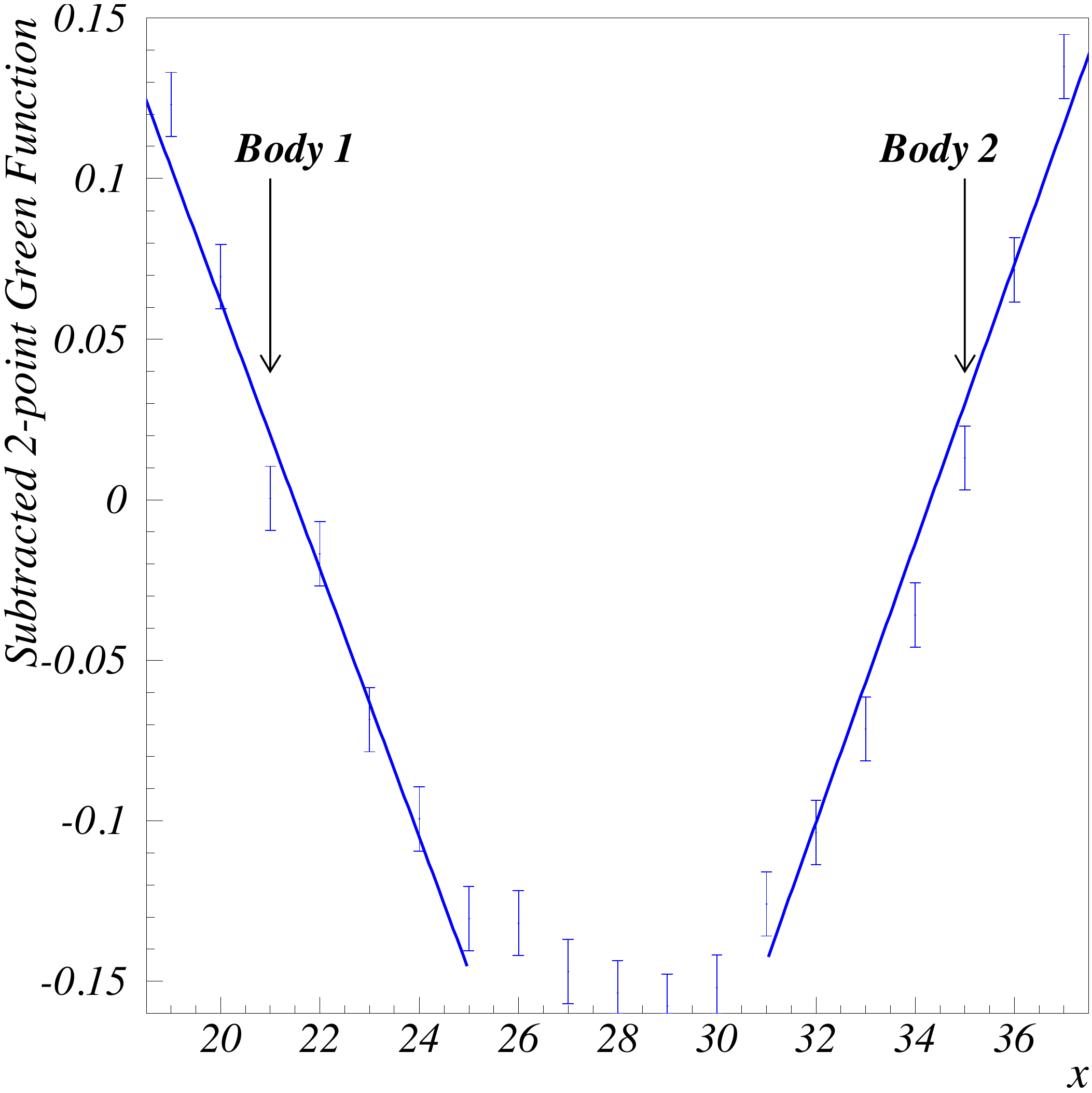}
\caption{\label{Flo:gflin}  
Potential around two massive bodies, with the $1/x$ (free--field, Newtonian case) contribution subtracted. 
The straight lines demonstrate the approximate linear behavior of the potential away from the 
mid-distance between the bodies  ($x=28$).
There, by symmetry, the potential must flatten, as the calculation indeed shows.
The potential was calculated in the static limit with Eq.~(\ref{eq:EH}) for $n \leq 2$.
The two sources are located on the $x-$axis
at $d=\pm7$ lattice spacings $u$ from the lattice center $x=28$, $y=0$
and $z=0$. The coupling is $16\pi GM=5.6\times10^{-5}~u$, the lattice size is $N=85$, the 
decorrelation parameter~\citep{Deur_DM-EPJC} is $N_{\mathrm{cor}}=20$ and 
$N_{\mathrm{s}}=3.5\times10^{4}$ decorrelated paths were used.
As boundary conditions, we used both random field values at the lattice edges, or Dirichlet 
boundary conditions. The resulting potentials are similar.}
\end{figure}

$G_{2p}(x_{1}-x_{2})$ is computed numerically on a 
cubic lattice of $N^{3}$ sites to which a field of value $\varphi$ is associated. 
The initial values of $\varphi$ at each site is chosen randomly.
The ensemble of the $N^{3}$ values is known as a field configuration. 
A physical configuration should be such that $S_s$ is minimized {\it{viz}}
the field verifies the Euler-Lagrange equations of motion.
To determine numerically these proper configurations, one must first perform a Wick rotation:
$\mathrm{e}^{\mathrm{-i}S_s/\hbar}\to\mathrm{e}^{-S_s/\hbar}$. Euclidean and
Minkowski actions being the same, $S_s$ remains unchanged. 
One then follows the Metropolis algorithm iteratively: 
$S_s$ is computed on each sites. The value of $\varphi$ at a given site is randomly varied 
and the consequent modification $\Delta S_s$ is calculated. If $\Delta S_s \leq 0$,
the new $\varphi$ value tends to minimize $S_s$. If so, one retains the new $\varphi$ 
value since the configuration is now closer to one obeying the equations of motion. 
If $\Delta S_s > 0$, one keeps the new $\varphi$ if $\mathrm{e}^{-\Delta S_s/\hbar} > \varepsilon$, 
with $\varepsilon$ randomly chosen between 0 and 1. Otherwise, the new $\varphi$ is rejected. 
As one iterates the procedure over all the sites, one converges to a configuration following 
the Euler-Lagrange equations, i.e.~the configuration probability distribution obeys
$\mathrm{e}^{-S_s/\hbar}$. 
This operation is repeated and the results averaged until they converge and until the statistical uncertainty 
inherent to the random method becomes small enough. Figure~\ref{Flo:gflin} shows an example
of calculation which resulted in a linear potential around two forces, \emph{viz} a constant force.

The path-integral formalism at the basis of the lattice approach produces intrinsically
quantum results. However, the results used in the present manuscript are classical because 
the lattice time is taken to infinity~\citep{Buchmuller:1997nw}, also known as the high-temperature limit.
This can be understood as follow: since the system is static, 
$S_{\mathrm{s}}\equiv{\int{\mathrm{d}^{4}x\,\mathcal{L}}=\tau S}$,
with $S\equiv{\int{\mathrm{d}^{3}x\,\mathcal{L}}}$ and $\tau=\intop_{t_{0}}^{\infty}{\mathrm{d}t\to\infty}$.
The exponential of Eq.~(\ref{eq:2pt green}) becomes 
$\mathrm{e}^{\mathrm{-i}S_{\mathrm{s}}/\hbar}=\mathrm{e}^{\mathrm{-i}\tau S/\hbar}$ and, just like 
$\hbar  \to 0$ suppresses quantum effects, the $\hbar/\tau \to 0$ when $\tau\to\infty$ yields the
classical limit. 

\begin{figure}
\center
\includegraphics[width=0.46\textwidth]{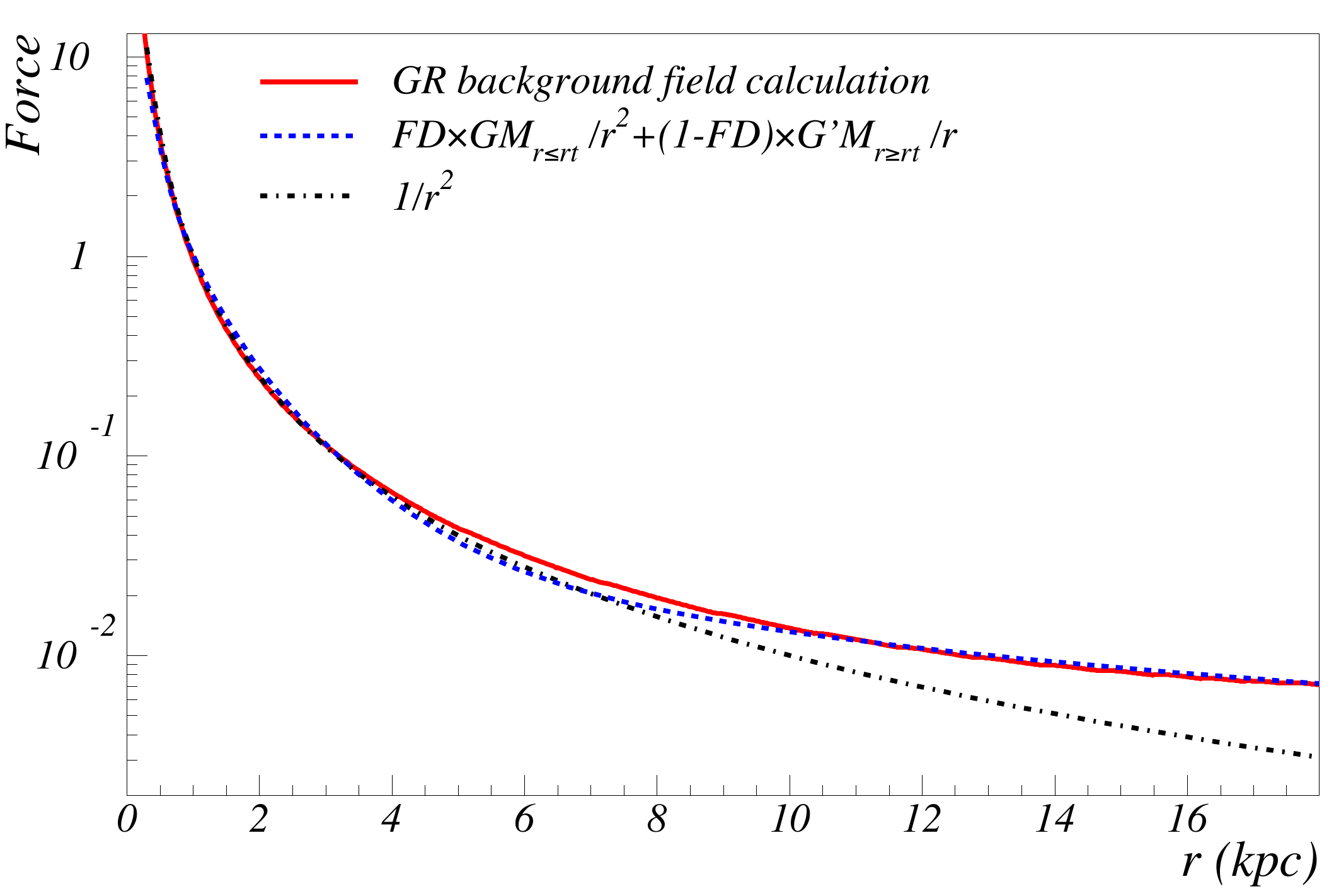} 
\caption{Distance dependence of the force obtained using a mean-field approximation to 
compute the self-interaction effects in a disk galaxy (solid red line)~\citep{deur:2020}. 
The total galaxy baryonic mass is $M_{\rm tot} = 5\times10^{11}$M$_\odot$ and has an exponentially 
decreasing density profile characterized by $h = 1.5~{\rm kpc}$. The dashed blue line is a parameterization 
of the force using the same method as for Model 1 described in the manuscript: 
below a transition scale $r_t=2h$, a Newtonian potential ($1/r^2$ force) is used and a 2-dimensional 
logarithmic potential ($1/r$ force) is used.  
A Fermi-Dirac function ($FD$) of width $r_t$ is used to smoothly connect the two domains.
$M_{r \leq r_t}$ is the mass enclosed within $r_t$ and $M_{r \geq r_t} = M_{\rm tot} - M_{r \leq r_t}$ is the 
mass outside $r_t$.
The dashed-dot black line is the expectation from a pure Newtonian potential.}
\label{fig:force_bkg}
\end{figure}

The method summarized in this appendix has been checked in different ways~\citep{Deur_DM-EPJC}:
\begin{itemize}[noitemsep,topsep=0pt]
\item  Analytically known potentials for free-field (i.e.~theories without self-interacting terms) have been 
recovered for both massive (Yukawa potential) or massless (Coulomb and Newtonian potentials) fields in 
three spatial dimensions. They were also satisfactorily verified in the two spatial dimensions case. 
\item The analytically known potential~\citep{Frasca:2009bc} for the self-interacting $\phi^4$ theory 
was retrieved.
\item The phenomenological static potential for the strong interaction 
(Cornell potential~\citep{Eichten:1974af}) was recovered once short distance quantum effects, {\it{viz}} the 
scale dependence of $\alpha_s$, were accounted for.
\item The logarithmic potential resulting from the lattice calculation extended from a 2-body system to a thin 
disk system was also obtained~\citep{deur:2020} by estimating GR's self-interaction effects in a typical disk 
galaxy using a mean-field method that is not based on Eq.~(\ref{eq:EH}), see Fig.~\ref{fig:force_bkg}.
\end{itemize}

\begin{figure}
\center
\includegraphics[width=0.46\textwidth,height=3in]{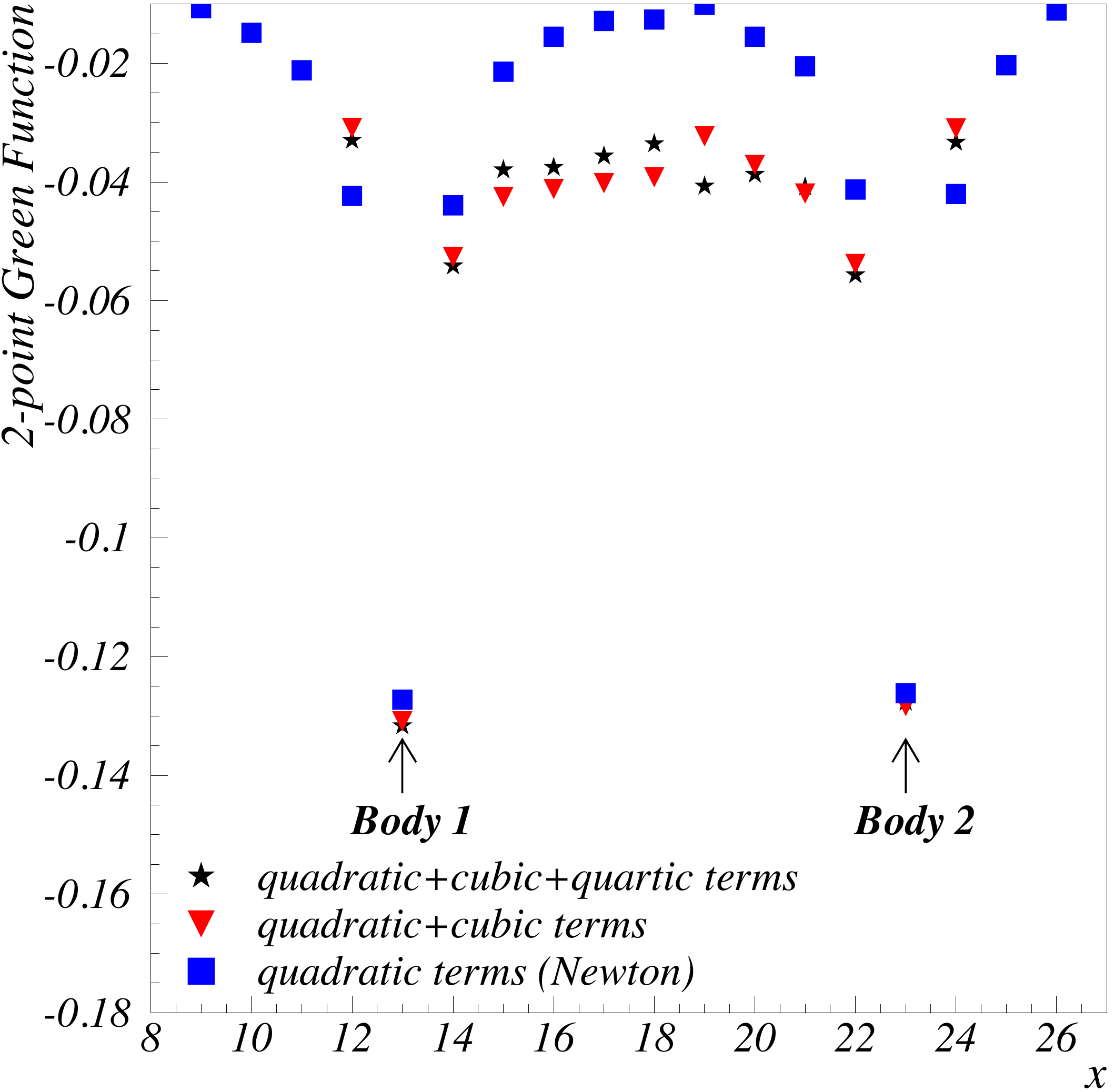} 
\includegraphics[width=0.46\textwidth,height=3in]{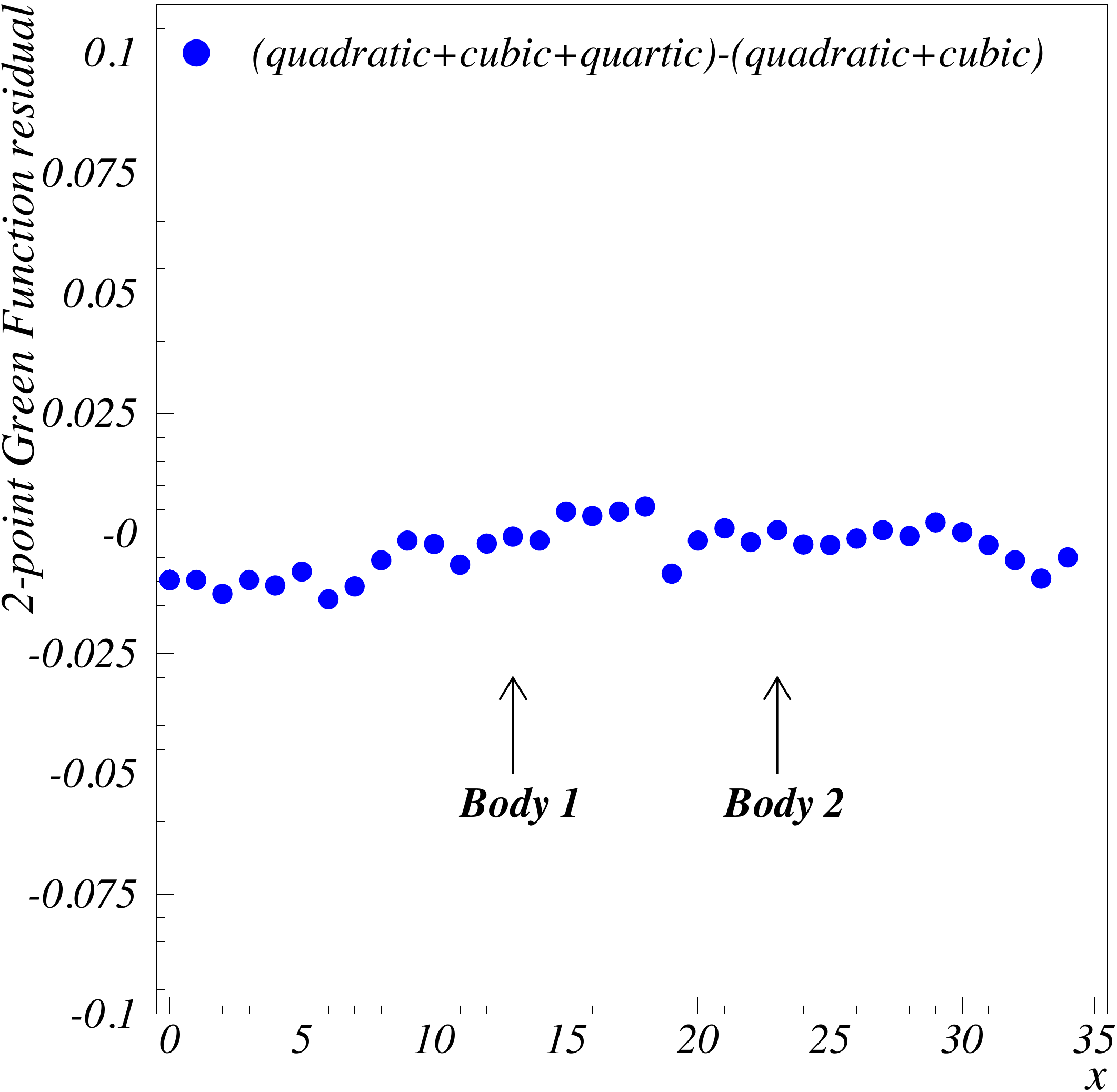} 
\caption{
Left: potential between two massive bodies, calculated in the static limit 
with Eq.~(\ref{eq:EH}) for $n \leq 2$ (black stars), $n \leq 1$ (red triangles), and $n=0$
(Newtonian case, blue squares). The two sources are located on the lattice $x-$axis
at $\pm5$ lattice spacings $u$ from the lattice center at $x=18$, $y=0$ and $z=0$. 
The coupling value is $16\pi GM=1\times10^{-4}~u$, the lattice size $N=35$, the decorrelation 
parameter~\citep{Deur_DM-EPJC} $10$ and $N_{\mathrm{s}}=3\times10^{4}$ decorrelated 
paths were used. Random field values at the lattice edges 
were used for boundary conditions (similar
results are obtained when using Dirichlet boundary conditions). 
With these calculation parameters, the difference between the cases $n \leq 2$ and $n \leq 1$
is typically less than 10\% of that between the $n=0$ and $n \leq 1$ cases. 
Right: residual between the potential calculated with $n \leq 2$ for Eq.~(\ref{eq:EH})
and the one calculated for $n \leq 1$, shown with the same vertical scale range as that of the left panel 
for easier comparison. This residual is about $\simeq 10^{-3}$, small compared to the $n \leq 1$ and 
$n \leq 2$  difference ($4\times 10^{-2}$ between the two bodies), and the potential scale ($0.13$). 
This justifies the truncation of Eq.~(\ref{eq:EH}) to $n=2$.
}
\label{fig:greenfunc_resid}
\end{figure}

All the GR lattice calculations were done with the Lagrangian given by Eq.~(\ref{eq:EH}) with 
$n=0$, $n \leq 1$ and $n \leq 2$. 
In the static limit, the ratio of two consecutive field terms $n$ and $n+1$ is $(16 \pi GM)^{1/2} \varphi_{00}$, with $16\pi GM$ $\ll 1$ suggesting that Eq.~(\ref{eq:EH}) can be truncated at low $n$.
The $n=0$ results, which reproduce the expected free-field 
potentials, differ significantly from the $n \leq 1$ and $n \leq 2$ 
results once the system mass $M$ is large enough (given the geometry of the system) so that
GR has entered its non-linear regime. However, the $n \leq 1$ and $n \leq 2$ calculations yielded similar 
results, see Fig.~\ref{fig:greenfunc_resid}. Thus, the first self-interaction term ($n=1$) dominates
and is enough to describe the effects of field self-interaction. For smaller values of $M$, 
the $n=0$ contribution to the potential dominates the $n>0$ contributions.

~
\begin{figure}
\center
\includegraphics[width=0.4\textwidth]{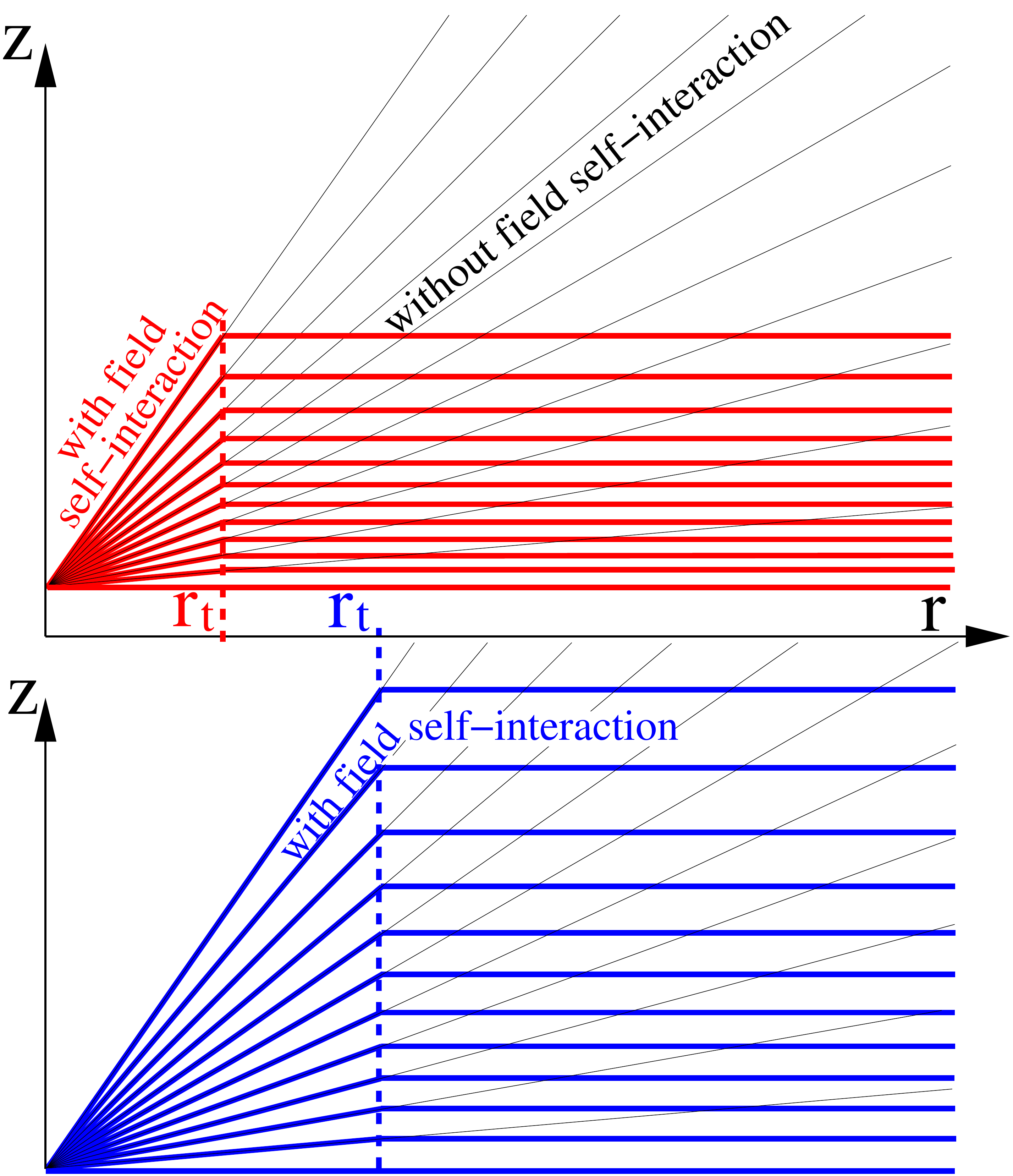} 
\caption{Dependence of $G'$ on the transition scale $r_t$. Field lines emerge radially for a source (here, for 
clarity, only those emerging from the galaxy center are shown). A coupling constant, here $G'$, determines
the density of the field lines emerging from the source (or, in this sketch, the number of field lines 
represented). Since the field possesses energy-momentum, it interacts gravitationally with itself and with 
masses. Field self-interactions and interactions of the field with the massive disk in the $z=0$ plane bend the 
field lines. At smaller $r$ where the field lines are still radially distributed, the force behaves as $1/r^2$ (3D 
regime). At larger $r$ where they are parallel to each others in a given vertical plane, but are still radially 
distributed in the disk plane---because of the cylindrical symmetry of the disk---the force behaves as $1/r$ (2D 
regime). For simplicity, the transition distance between the two regimes is shown here to be infinitely short. 
For a small $r_t$ (top panel), the field lines at large $r$ are denser. For a larger $r_t$ (bottom panel), the field 
lines are sparser. Their density is approximately proportional to $r_t$. Since a force coupling constant reflects 
the overall density of its field lines (i.e.~ignoring the $r$-dependence) the coupling in the 2D case 
approximately obeys $G' = G/r_t$ assumed in this article.}
\label{field_line_sketch}
\end{figure}

\section*{Appendix C: Non-universality of $G'$ and its value for infinitely thin disks}

The expression of the gravitational force confined in 2D is $G'Mm/r$. Therefore, one would naturally expect 
$G'$ to be universal, like $G$ in the 3D case. Furthermore, in the analogous QCD case, the effective coupling 
$\sigma$ (the analog of $G'$), known as the QCD string tension, is indeed universal with a value of $0.18$ 
GeV$^2$~\citep{Deur:2016tte}. However, $G'$ is not universal, but depends on the geometry of the galaxy, its 
mass, and its density distribution.

To understand why, it is convenient to visualize a force as a field flux through an elementary surface. The 
force coupling constant controls the overall density of the field lines for a unit of charge or mass. Its value 
does not change the $r$-dependence of the force\footnote{This is true only in the classical case. Running 
couplings in quantum field theory do affect the $r$-dependence because short distance quantum effects are 
folded into the definition of the coupling~\citep{Deur:2016tte}. This definition of the coupling at quantum scale 
is conventional and, in any case, irrelevant here.}. Likewise, $G'$ determines the overall density of field lines 
passing through an elementary segment. This density depends on how early the transition from 3D to 2D 
occurs, as sketched in Fig.~\ref{field_line_sketch}, where, for clarity, we have drawn only the field lines 
emerging from the center of the galaxy, its densest locus. The transition occurs early for large disk densities, 
or can be delayed by the presence of a spherically symmetric bulge. Therefore, $G'$ depends on both the 
morphology and mass distribution of the galaxy components. In the case of an early transition (red lines in 
Fig.~\ref{field_line_sketch}), the field lines are denser and $G'$ is large. For a later transition (blue lines), the 
field lines are sparser and $G'$ is smaller. Thus, $G'$ is not universal and approximately obeys $G' = G/r_t$.

In the QCD case, $\sigma$ is universal because there is no geometrical or color charge variation: for the 
heavy meson case to which $\sigma$ applies, two static pointlike sources of unit color charge are 
invariably considered, with the flavors of the sources and their type of color having no influence on the 
force. Therefore the same distortion of field lines occurs, regardless of the type of meson considered, 
and $\sigma$ is universal. 

One may also ask what is the value of $G'$ for a pure (bulge-less) disk, since there is no 
bulge-to-disk transition.
Inside the disk, the mass distribution is approximately isotropic so the scale height $h_z$ of the disk 
sets a first limit for the scale: one expects $r_t \propto h_z$. 
However, considering an infinitely thin disk reveals that a transition scale $r_t$ emerges dynamically, which 
may be larger that $h_z$. Even for an infinitely thin disk ($h_z=0$), it takes a length $r_t$ for the initially 
radially distributed field lines to bend into parallel field lines. The mass and its distribution thus determine 
$r_t$: the larger the mass and the more concentrated the density, the smaller $r_t$. The dynamical 
emergence of $r_t$ in a massive infinitely thin disk is  analogous to the emergence of the confinement scale 
of QCD, or to the energy difference arising between the ground state and first exited levels in atoms or more 
complex materials in atomic or solid state physics, {\it{viz}} computing $r_t$ is a spectral gap problem. The 
gap problem is notoriously difficult~\citep{gappb} and without known analytical solution.
Therefore, even for infinitely thin disks, $r_t$---or equivalently $G'$---is non-universal and cannot presently be 
analytically calculated from first principles. It can be obtained from numerical calculations such as those in 
Refs.~\citep{Deur_DM-PLB, Deur_DM-EPJC}, or assessed phenomenologically as done in this article.

\bibliography{dst_2020}{}
\bibliographystyle{aasjournal}



\end{document}